\documentclass[twocolumn]{aastex631}

\shorttitle{Nuclear Transients in ZTF}
\shortauthors{Dgany et al.}

\graphicspath{{./}}
\usepackage{enumitem}

\begin{document}

\title{Needle in a Haystack: Finding Supermassive Black Hole-Related Flares in the Zwicky Transient Facility Public Survey}

\author[0000-0002-7579-1105]{Yael Dgany}
\affiliation{The School of Physics and Astronomy, Tel Aviv University, Tel Aviv 69978, Israel}

\author[0000-0001-7090-4898]{Iair Arcavi}
\affiliation{The School of Physics and Astronomy, Tel Aviv University, Tel Aviv 69978, Israel}
\affiliation{CIFAR Azrieli Global Scholars program, CIFAR, Toronto, Canada}

\author[0000-0002-7466-4868]{Lydia Makrygianni}
\affiliation{The School of Physics and Astronomy, Tel Aviv University, Tel Aviv 69978, Israel}

\author[0000-0002-7472-1279]{Craig Pellegrino}
\affiliation{Las Cumbres Observatory, 6740 Cortona Drive, Suite 102, Goleta, CA 93117-5575, USA}
\affiliation{Department of Physics, University of California, Santa Barbara, CA 93106-9530, USA}

\author[0000-0003-4253-656X]{D. Andrew Howell}
\affiliation{Las Cumbres Observatory, 6740 Cortona Drive, Suite 102, Goleta, CA 93117-5575, USA}
\affiliation{Department of Physics, University of California, Santa Barbara, CA 93106-9530, USA}

\correspondingauthor{Iair Arcavi}
\email{arcavi@tauex.tau.ac.il}

\begin{abstract}
Transient accretion events onto supermassive black holes (SMBHs), such as tidal disruption events (TDEs), Bowen Fluorescence Flares (BFFs), and active galactic nuclei (AGNs), which are accompanied by sudden increases of activity, offer a new window onto the SMBH population, accretion physics, and stellar dynamics in galaxy centers. However, such transients are rare and finding them in wide-field transient surveys is challenging. Here we present the results of a systematic real-time search for SMBH-related transients in Zwicky Transient Facility (ZTF) public alerts, using various search queries. We examined 345 rising events coincident with a galaxy nucleus, with no history of previous activity, of which 223 were spectroscopically classified. Of those, five (2.2\%) were TDEs, one (0.5\%) was a BFF, and two (0.9\%) were AGN flares. Limiting the search to blue events, the fraction of TDEs nearly doubles to 4.1\%, and no TDEs are missed. Limiting the search further to candidate post-starburst galaxies increases the relative number of TDEs to 16.7\%, but the absolute numbers in such a search are small. The main contamination source is supernovae (95.1\% of classified events), of which the majority (82.2\% of supernovae) are of Type Ia. In a comparison set of 39 events with limited photometric history, the AGN contamination increases to $\sim$30\%. Host galaxy offset is not a significant discriminant of TDEs in current ZTF data, but might be useful in higher-resolution data. Our results can be used to quantify the efficiency of various SMBH-related transient search strategies in optical surveys such as ZTF and the Legacy Survey of Space and Time.

\end{abstract}


\section{Introduction}

Wide-field optical surveys have recently found new types of transients occurring exclusively in galaxy centers. These transients are thought to be associated with enhanced accretion events onto supermassive black holes (SMBHs). As such, they have the potential to reveal the presence and properties of otherwise inactive SMBHs, as well as constrain physics of accretion and related radiative processes. Notable examples of such transients are optical-ultraviolet tidal disruption events (TDEs) and Bowen Fluorescence Flares (BFFs). Both types of events are characterized by a sudden increase of flux by several orders of magnitude and are thus much more dramatic than the few tens of percent level variability seen in most active galactic nuclei (AGNs), which host SMBHs with steadily accreting material.

A TDE is the result of the disruption of a star by an SMBH \citep{Hills1975}. In such an event, half of the stellar material is expected to accrete onto the SMBH \citep{Rees1988}. For disruptions occurring outside the event horizon (expected for solar-type stars disrupted by SMBHs of masses $\lesssim10^8\,\textrm{M}_{\odot}$, for example), the accretion event will be accompanied by an observable flare. Several such flares have been detected in X-rays, as expected from directly observed accretion emission (see \citealt{Saxton2020} for a recent review). However, somewhat surprisingly, a class of optical-ultraviolet TDEs has also been discovered \citep{Gezari2012,Arcavi2014}. These events are mostly blue, with blackbody temperatures of a few $10^4$\,K lasting for several months to years, and show broad emission lines of H and/or He in their spectra. The emission mechanism leading to these observed properties is a topic of active debate (see \citealt{vanVelzen2020} and \citealt{Gezari2021} for recent reviews). 

In addition to their emission mechanism puzzle, optical-ultraviolet TDEs show a peculiar and strong host galaxy preference for post-starburst galaxies \citep{Arcavi2014,French2016,French2020}. This preference is not yet fully understood, but might be related to the spatial distribution and dynamics of various stellar populations in the centers of such galaxies \citep{French2020hst}. Studying optical-ultraviolet TDEs can thus also help shed light on the stellar dynamics in galaxy nuclei, which are responsible for driving up TDE rates in post-starburst environments \citep[e.g.][]{Madigan2018}. 

Like TDEs, BFFs are also blue and show H and He lines in their spectra, leading \cite{Tadhunter2017} to classify the first such observed event as a TDE. However, a second \citep{Gromadzki2019} and then third \citep{Trakhtenbrot2019} event showed that their typical spectral line widths are much narrower than those of TDEs, and that their light curves decline much more slowly than those of TDEs. This led \cite{Trakhtenbrot2019} to classify BFFs as a separate observational class. There are now also hints that BFFs occur in previously existing AGN \citep{Makrygianni2023}, meaning that they could be the result of accretion instabilities in an AGN disk, or of a TDE occurring in an AGN and interacting with its existing accretion disk \citep[e.g.][]{Chan2019}.

BFFs are named as such because they exhibit certain emission lines (such as He\,II 4686\AA\, and N\,III 4640\AA, among others) that are associated with the Bowen fluorescence mechanism \citep{Bowen1928}. In this mechanism, extreme-ultraviolet and X-ray photons excite certain He II transitions, which in turn launch a cascade of transitions observed in the optical and ultraviolet regimes. This process requires the presence of extreme-ultraviolet photons hitting high-density and high-optical-depth material, and is was indeed predicted decades ago to occur in AGNs \citep{Netzer1985}. Since the identification of this mechanism in BFFs, it has also been suggested to occur in some TDEs \citep{Blagorodnova2019,Leloudas2019}, hinting at a possible connection between the conditions of matter and radiation in these two types of events related to SMBH accretion.

It is clear that studying more TDEs and BFFs is necessary in order to better constrain their nature, emission mechanisms, and the physics they can teach us in relation to SMBHs and their associated accretion processes. However, these events are intrinsically rare. The exact TDE rate remains uncertain, but is likely to be in the range of $10^{-5}$--$10^{-4}$ events per galaxy per year \citep[e.g.][]{Wang2004,Stone2016}. The BFF rate is not yet estimated at all, but observationally, they are less common than TDEs (this could be due in large part to selection effects, as discussed below). In addition to their intrinsic rarity, finding TDEs and BFFs is also observationally challenging. As events that occur in galaxy centers, their detection is contaminated by image-subtraction artifacts, ``regular'' AGN activity, unresolved non-SMBH-related transients, and even variable stars that cannot be easily distinguished from distant or compact galaxies. For these reasons, only a few dozen TDEs and a few BFFs have been identified so far.

Attempts have been made to devise selection criteria to weed out such transients from the large alert streams produced by wide-field transient surveys. Such criteria typically include selecting candidates by the significance of the flare (since TDEs and BFFs are luminous), color (since TDEs and BFFs are blue), and host properties (since TDE hosts are mostly quiescent). 

\cite{Hung2018} searched a set of 493 nuclear transients (0\farcs8 from their host galaxy center) from the intermediate Palomar Transient Factory, for events with $g-R<0$ mag residing in galaxies with $u-g>1$ mag and $g-r>0.5$ mag. These cuts reduced the set of candidates to just 26, of which two are TDEs. Still, the contamination fraction is large. A substantial amount of telescope time is required to vet 13 candidates (through spectroscopy or ultraviolet colors) for each bona fide TDE.

One way to further reduce the amount of transient contamination in TDE searches is to focus the search on galaxies most similar to the post-starburst hosts that TDEs seem to prefer. \cite{French2018} used galaxies from the Sloan Digital Sky Survey \citep[SDSS;][]{York2000} Data Release 12 main galaxy survey \citep{Strauss2002,Alam2015}, with similar spectral properties as those of actual TDE hosts, to train a machine-learning algorithm to identify such galaxies from photometry alone. They then used this algorithm to identify several tens of thousands of TDE host galaxy candidates in archival survey data. \cite{Arcavi2022} found indeed that using the \cite{French2018} catalog of galaxies reduces contamination by roughly a factor of 3-50 (depending on the subset of galaxies used from the catalog) compared to filtering just on quiescent galaxies, and that the only contaminant transients in such galaxies are Type Ia supernovae (SNe). That study, however, was based on archival data alone.

Here we perform a systematic real-time search for TDEs, BFFs, and other possible SMBH flares in the Zwicky Transient Facility \citep[ZTF;][]{Bellm2014,Graham2019} alert stream, as parsed by the Lasair broker \citep{Smith2019}. We use various search criteria that rely on candidate brightness and color and their host galaxy properties, and compare their effectiveness in selecting TDEs and BFFs against actual spectroscopic classifications obtained by us and by the rest of the community. We focus on rising events (i.e. events discovered before their peak), selected using visual inspection, for two main reasons. First, such events are more scientifically valuable, as they include the peak time and brightness, as well as the early pre-peak emission, both of which contain important information for constraining models of SNe and TDEs. In addition, rising events present a way of decreasing the number of events to a more manageable subsample for spectroscopic classification, while avoiding biasing the sample toward a particular class (the selection of rising events is done before their classifications are known).\footnote{One exception is that we would be less likely to catch rapidly rising events during their rise compared to slower-rising events, but the typical ZTF public survey cadence is $\lesssim3$ days, which should be enough to catch most rapidly evolving transients \citep[e.g.][]{Ho2021}.}

Our goal is to quantify the contamination fraction for the various search criteria and to check whether any of them miss TDEs and BFFs. Here, we do not constrain the intrinsic rates of SNe, TDEs, or BFFs in nature, but rather the observed fractions of events, to help guide searches in ongoing and future transient surveys and to help prioritize limited spectroscopic classification resources. We detail our search criteria in Section \ref{sec:methods}, present and analyze our results in Section \ref{sec:results}, and discuss them and conclude in Section \ref{sec:conclusions}.

\section{Methods}\label{sec:methods}

We searched the ZTF real-time alert stream for transients in galaxy centers every day between 2020 November 3 and 2022 March 6 (UT dates), with the exception of a $\sim$2-month break due to a ZTF technical outage between 2021 December 5 and 2022 February 17. In total, our search includes alerts from 414 days. We used the custom query builder on version 1.0 of the Lasair broker\footnote{\url{https://lasair.lsst.ac.uk}} to filter the alerts. Lasair uses a contextual classifier called Sherlock\footnote{\url{https://lasair.readthedocs.io/en/develop/core_functions/sherlock.html}}. Sherlock is a boosted decision tree algorithm that provides an initial classification of every nonmoving object by performing a spatial crossmatch against data from historical and ongoing astronomical surveys, including catalogs of nearby galaxies, variable stars, and AGNs (see Section 4.2 of \citealt{Smith2020} for more details).

Our queries, which are based on the TDE queries by M. Nicholl on version 1.0 of Lasair\footnote{\url{https://lasair.roe.ac.uk/filters/94/} and \url{https://lasair.roe.ac.uk/filters/95/}}, filter ZTF alerts according to the following criteria (for each, we state the corresponding Lasair query condition):
\begin{enumerate}
    \item The candidate is within a certain threshold distance of the nearest Sherlock catalog source. For 80\% of our sample, we choose a threshold of 0\farcs5\footnote{This value was chosen given that the ZTF pixel scale of 1\arcsec\ per pixel results in a typical centroiding accuracy of $\lesssim$0\farcs3, which we increase to 0\farcs5 to be inclusive}. For the rest, we increased the separation threshold to 1\arcsec\ to check if this has a strong effect on the results:\\
    {\tt sherlock\_classifications.separationArcsec < 0.5}\\
    or\\
    {\tt sherlock\_classifications.separationArcsec < 1}\\
    We found that the value of the threshold has no significant effect on the results (see Appendix B), and therefore analyzed the joint sample of both separation thresholds together to increase our sample size. This condition, regardless of the separation threshold, filters out ``hostless'' events, i.e. those with no host in the Sherlock catalog.
    
    \item The nearest catalog source is likely a galaxy rather than a star\footnote{{\tt sgscore1} is based on a random forest classifier trained and implemented in the ZTF alerts by \cite{Tachibana2018}. An {\tt sgscore} value closer to 0 means that the nearest source in the Panoramic Survey Telescope and Rapid Response System \citep[Pan-STARRS;][]{Kaiser2010} first survey \citep[PS1;][]{Chambers2016} catalog is more likely a galaxy, while a value closer to 1 means it is more likely a star.}:\\
    {\tt objects.sgscore1 < 0.5}
    
    \item The Sherlock classification of the candidate is either ``SN'' (Supernova) or ``NT'' (Nuclear Transient):\footnote{The other Sherlock classifications, which we exclude from our search, are: ``VS'' (Variable Star), ``CV'' (Cataclysmic Variable), ``BS'' (Bright Star), ``AGN'', ``Orphan'' (if the transient fails to match against any cataloged source), and ``Unclear''.}\\ 
    {\tt sherlock\_classifications.classification in (`SN',`NT')} 
    
    \item The candidate does not have detections more than 100 days ago (indicating that it might be a variable, rather than a transient source, though some past detections could be artifacts):\footnote{A ZTF alert is reported to the brokers with a 30 day history, which may contain prediscovery detections. Lasair marks the time of the first detection in this 30 day history as {\tt jdmin}.}\\
    {\tt objects.jdmin > JDNOW()-100}
    
    \item The candidate does not have a ZTF17 or ZTF19 name, meaning that it was not created by ZTF in 2017 or 2019 (this is another way of filtering out variable sources):\footnote{Sporadic false detections in galaxy centers may occur, sometimes years before a real event occurs at the same position, and based on false detections that are later filtered out by the brokers. In such a case, the real event would have an old name from when the false detection occurred years before. Removing such events might thus undesirably filter out interesting candidates. In order to avoid losing many candidates, but still not being inundated with variable sources, we decided to allow events with ZTF18 names (several bad subtractions in 2018 caused false events then; E. Bellm, private communication), while removing those with ZTF17 and ZTF19 names.}\\
    {\tt objects.objectId NOT LIKE `ZTF17\%' AND objects.objectId NOT LIKE `ZTF19\%'}
    
    \item The candidate is not a previously classified SN:\\
    {\tt crossmatch\_tns.tns\_prefix != `SN'}

    \item The candidate has $<$3 of its detections deemed unreliable (i.e. which are not marked as good quality and/or the candidate is dimmer than the reference):\footnote{{\tt ncand} is the total number of detections from ZTF, which can be either positive or negative subtraction residuals (i.e. a brightening or fading with respect to the reference image). {\tt ncandgp} counts only `good and positive' detections, i.e. a positive flux with respect to the reference and having a ZTF machine-learning real-bogus score $>$0.75. This criterion requires that most detections are good and positive, but allows for one or two light-curve points with poor real-bogus scores if, for example, the transient was detected when it was very young and the subtraction residuals at the earliest epochs have a low real-bogus score due to a relatively low signal-to-noise ratio.}\\
    {\tt objects.ncand - objects.ncandgp < 3}
    
    \item At least one of those detections was no more than 14 days ago (in order to avoid old objects that might already be fading):\\
    {\tt objects.ncandgp\_14 > 1}
    
    \item The candidate is more than 10$^\circ$ away from the Galactic plane (in order to filter stellar flares or variability):\\
    {\tt objects.glatmean > 10 OR objects.glatmean < - 10}
    
\end{enumerate}

Conditions 4 and 5 could introduce a bias against finding BFFs \citep[which might be associated with preexisting AGNs;][]{Makrygianni2023} and TDEs occurring in AGNs. However, these conditions are necessary in order to remove ``normal'' AGN activity, which can otherwise be a major contaminant (see below).

In addition to conditions 1--9, we create two variations of the query, each with a different magnitude limit:

\begin{enumerate}[label=10\alph*.]
    \item The latest $g$- or $r$-band magnitude of the candidate is brighter than 19:\\
    {\tt objects.rmag < 19 OR objects.gmag < 19}

    \item The latest $g$- or $r$-band magnitude of the candidate is brighter than 19.5:\\
    {\tt objects.rmag < 19.5 OR objects.gmag < 19.5}
\end{enumerate}

The motivation for these variations is due to several spectrographs on dynamically scheduled telescopes (which are ideal for the rapid classification of transients) - such as the Floyds spectrographs \citep{Brown2013} on the Las Cumbres Observatory Faulkes Telescopes North (FTN) and South (FTS) and the SPectrograph for the Rapid Acquisition of Transients \citep[SPRAT;][]{Piascik2014} on the Liverpool Telescope - being on 2 m class telescopes, with a typical limiting magnitude of 19. Similarly, the advanced extended Public European Southern Observatory (ESO) Spectroscopic Survey for Transient Objects \citep[ePESSTO+, a continuation of PESSTO;][]{Smartt2013}, responsible for a large number of transient classifications, uses the ESO Faint Object Spectrograph and Camera v2 \citep[EFOSC2;][]{Buzzoni1984} on the 3.6 m New Technology Telescope (NTT) to reach a magnitude of 19.5. 

For each of these variations, we create three subvariations: one without any additional conditions, one with an additional condition on the color of the event:
\begin{enumerate}[label=11.]
    \item The candidate has a $g-r$ magnitude difference $<$0.05 (in order to select only blue events, since TDEs are observed to maintain roughly constant blue colors for months; see e.g. \citealt{vanVelzen2020}):\footnote{The $g-r$ color is evaluated on the most recent night with positive detections (relative to the reference) in both bands. We choose a threshold of 0.05, to be slightly more conservative than the threshold of 0 suggested by \cite{Hung2018}. We do not take into account magnitude errors here.}\\ 
    {\tt objects.g\_minus\_r < 0.05}
\end{enumerate}
and, finally, one subvariation that searches for candidates coincident with galaxies from the \cite{French2018} catalog of likely TDE hosts \citep[hereafter referred to as post-starburst, or PS, galaxies;][]{Arcavi2014,French2016}, as implemented in the ``E+A Galaxies'' watchlist\footnote{\url{https://lasair.roe.ac.uk/watchlist/321/}} on Lasair. 

In total we have six queries, which we hereby number as follows:

\begin{enumerate}[label=\Roman*.]
    \item Conditions 1--9, with a limiting magnitude $<$19 (Condition 10a), blue (Condition 11), and in a PS galaxy.
    \item Conditions 1--9, with a limiting magnitude $<$19 (Condition 10a) and blue (Condition 11).
    \item Conditions 1--9, with a limiting magnitude $<$19 (Condition 10a).
    \item Conditions 1--9, with a limiting magnitude $<$19.5 (Condition 10b), blue (Condition 11), and in a PS galaxy.
    \item Conditions 1--9, with a limiting magnitude $<$19.5 (Condition 10b) and blue (Condition 11).
    \item Conditions 1--9, with a limiting magnitude $<$19.5 (Condition 10b).
\end{enumerate}

Obviously, these are not independent, with some queries being subsets of others, and all being subsets of Query VI. These queries produced roughly 30 new candidates per day in total, which we inspected manually. Only those showing a coherently rising light curve were marked as candidates of interest. Candidates not obviously rising at discovery were monitored for an extra epoch of ZTF photometry and checked again. Candidates for which it was still not clear whether they were rising or not were monitored for another week. This step removed events that had a flat, varying, or incoherent light curve, which could be due to ``normal'' AGN variability, the stellar variability of Galactic objects, or artifacts of the ZTF image-subtraction pipeline. In addition, this removed true transients that were already after their peak luminosity and that are not part of our sample as defined here. Of all our filtering steps, this is the most subjective, as it requires visual inspection, rather than some strict criterion for what constitutes a ``coherently rising'' light curve. However, by checking each candidate during multiple epochs, we aim to make this step as inclusive as possible. In addition, since this step is performed before the classification of the candidate is known, it should not bias the search against a particular type of transient (except extremely rapidly rising events, with rise times $\lesssim$3 days). 

After these cuts, we were left with a total of 345 candidates of interest (from our entire 414 day search, i.e. $\sim$0.83 candidates of interest per day, on average), which we attempted to classify spectroscopically within a few days of discovery. 

Version 3.0 of Lasair (also known as ``Iris'') was released in 2021 March. To improve performance, not all of the information that was available in version 1.0 (such as the full detection histories of all candidate events) was carried over to version 3.0. To check for any differences in query results, we add four more queries that we ran on Lasair 3.0 (Iris) between 2022 April 6 and 2022 August 2 (for a total of 118 days):

\begin{enumerate}[label=\Roman*.]\setcounter{enumi}{6} 
    \item Conditions 1--9, with a limiting magnitude $<$19 (Condition 10a).
    \item Conditions 1--9, with a limiting magnitude $<$19.5 (Condition 10b).
    \item Conditions 1--9, with a limiting magnitude $<$19 (Condition 10a) and in a PS galaxy.
    \item Conditions 1--9, with a limiting magnitude $<$19.5 (Condition 10b) and in a PS galaxy.
\end{enumerate}

Color information was not available as a query parameter in Iris, therefore here we cannot filter by condition 11 here. We perform the same manual cuts as above and are left with 39 events, which is an average of 0.33 candidates per day. Version 4.0 of Lasair was released in 2022 May, but we do not test it here.   

We obtain a total of 345 candidates of interest from Lasair 1.0 (310 of which from using a separation threshold of 0\farcs5 in Criterion 1, with the rest from using a separation threshold of 1\arcsec) and 39 candidates of interest from Lasair 3.0 (all of which from using a separation threshold of 1\arcsec).

For all those brighter than 19th magnitude, we requested spectra through the Las Cumbres Observatory Floyds spectrographs mounted on the 2 m FTN and FTS telescopes at Haleakala (United States) and Siding Spring (Australia) observatories, respectively. Weather, technical issues, and oversubscription of the telescopes mean that not all the requested spectra were obtained or that some were obtained first by the community and reported to the Transient Name Server (TNS)\footnote{\url{http://www.wis-tns.org}}. We were able to obtain spectra of 83 candidates of interest, taken through a 2\arcsec\ slit placed on the candidate along the parallactic angle \citep{Filippenko1982}. One-dimensional spectra were extracted and the flux and wavelength were calibrated using the \texttt{floyds\_pipeline}\footnote{\url{https://github.com/LCOGT/floyds_pipeline}} \citep{Valenti2013}. Fainter targets, accessible from La Silla Observatory, were sent for consideration to the ePESSTO+ collaboration, for classification with the NTT. 

All of our classification spectra, as well as those obtained by the ePESSTO+ collaboration, were publicly reported to the TNS. Many of our candidates of interest were classified by other members of the community and also reported to the TNS. In total, 246 of our 384 candidates of interest (64.1\%) were classified on the TNS. We take these classifications as reported to the TNS and analyze their distribution in the next section.

\begin{figure}
\centering
\includegraphics[width=\columnwidth]{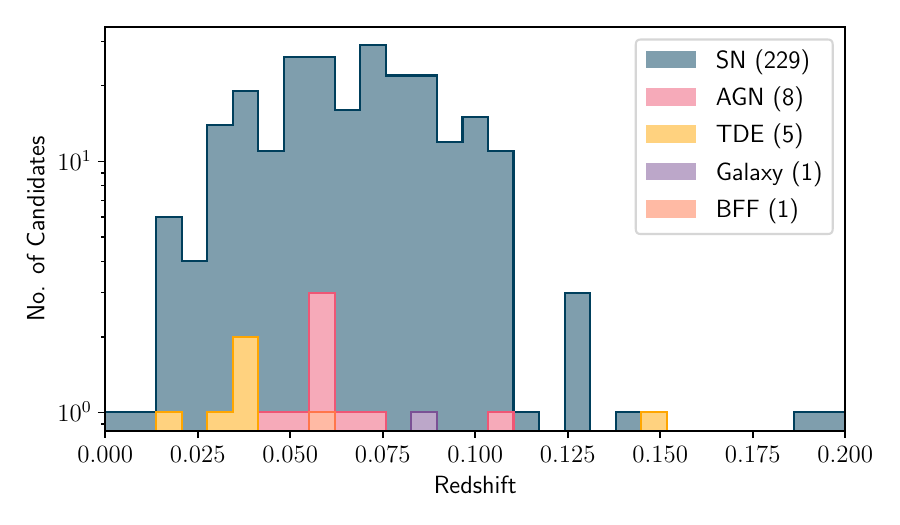}
\caption{\label{fig:redshifts}Redshift distribution (shown in a stacked histogram) of the 241 classified candidates of interest with redshifts on the TNS. The number of events in each class is denoted in parentheses in the legend.} 
\end{figure}

\section{Results and Analysis}\label{sec:results}

\begin{deluxetable*}{llllllllll}
\tablecaption{\label{tab:stats}Numbers and fractions of the classes of candidates of interest from the different queries.}
\tablehead{
\colhead{Query} & \colhead{Total} & \colhead{Not} & \colhead{SN} & \colhead{AGN} & \colhead{TDE} & \colhead{Other} & \colhead{Galaxy} & \colhead{BFF} & \colhead{Varstar}\\
\colhead{} & \colhead{Transients} & \colhead{Classified} & \colhead{} & \colhead{} & \colhead{} & \colhead{} & \colhead{} & \colhead{} & \colhead{}}
\startdata
\multicolumn{10}{c}{Lasair 1.0} \\
\hline
I: $<$19 Mag, Blue, and in PS & 6 & 1 & 4 & 0 & 1 & 0 & 0 & 0 & 0 \\
Percentage of All Transients &  & 16.67\% & 66.67\% & 0 & 16.67\% & 0 & 0 & 0 & 0 \\
Percentage of Classified Transients &  &  & 80.00\% & 0 & 20.00\% & 0 & 0 & 0 & 0 \\
\hline
II: $<$19 Mag and Blue & 116 & 19 & 91 & 0 & 5 & 0 & 1 & 0 & 0 \\
Percentage of All Transients &  & 16.38\% & 78.45\% & 0 & 4.31\% & 0 & 0.86\% & 0 & 0 \\
Percentage of Classified Transients &  &  & 93.81\% & 0 & 5.15\% & 0 & 1.03\% & 0 & 0 \\
\hline
III: $<$19 Mag & 213 & 41 & 163 & 2 & 5 & 1 & 1 & 0 & 0 \\
Percentage of All Transients &  & 19.25\% & 76.53\% & 0.94\% & 2.35\% & 0.47\% & 0.47\% & 0 & 0 \\
Percentage of Classified Transients &  &  & 94.77\% & 1.16\% & 2.91\% & 0.58\% & 0.58\% & 0 & 0 \\
\hline
IV: $<$19.5 Mag, Blue, and in PS & 9 & 3 & 5 & 0 & 1 & 0 & 0 & 0 & 0 \\
Percentage of All Transients &  & 33.33\% & 55.56\% & 0 & 11.11\% & 0 & 0 & 0 & 0 \\
Percentage of Classified Transients &  &  & 83.33\% & 0 & 16.67\% & 0 & 0 & 0 & 0 \\
\hline
V: $<$19.5 Mag and Blue & 193 & 71 & 116 & 0 & 5 & 0 & 1 & 0 & 0 \\
Percentage of All Transients &  & 36.79\% & 60.10\% & 0 & 2.59\% & 0 & 0.52\% & 0 & 0 \\
Percentage of Classified Transients &  &  & 95.08\% & 0 & 4.10\% & 0 & 0.82\% & 0 & 0 \\
\hline
VI: $<$19.5 Mag & 345 & 121 & 213 & 2 & 5 & 1 & 1 & 1 & 1 \\
Percentage of All Transients &  & 35.07\% & 61.74\% & 0.58\% & 1.45\% & 0.29\% & 0.29\% & 0.29\% & 0.29\% \\
Percentage of Classified Transients &  &  & 95.09\% & 0.89\% & 2.23\% & 0.45\% & 0.45\% & 0.45\% & 0.45\% \\
\hline
\multicolumn{10}{c}{Lasair 3.0 (Iris)} \\
\hline
VII: $<$19 Mag in Iris & 33 & 13 & 14 & 6 & 0 & 0 & 0 & 0 & 0 \\
Percentage of All Transients &  & 39.39\% & 42.42\% & 18.18\% & 0 & 0 & 0 & 0 & 0 \\
Percentage of Classified Transients &  &  & 70.00\% & 30.00\% & 0 & 0 & 0 & 0 & 0 \\
\hline
VIII: $<$19.5 Mag in Iris & 39 & 17 & 16 & 6 & 0 & 0 & 0 & 0 & 0 \\
Percentage of All Transients &  & 43.59\% & 41.03\% & 15.38\% & 0 & 0 & 0 & 0 & 0 \\
Percentage of Classified Transients &  &  & 72.73\% & 27.27\% & 0 & 0 & 0 & 0 & 0 \\
\hline
IX: $<$19 Mag, in Iris, and in PS & 1 & 0 & 0 & 1 & 0 & 0 & 0 & 0 & 0 \\
Percentage of All Transients &  & 0 & 0 & 100.00\% & 0 & 0 & 0 & 0 & 0 \\
Percentage of Classified Transients &  &  & 0 & 100.00\% & 0 & 0 & 0 & 0 & 0 \\
\hline
X: $<$19.5 Mag, in Iris, and in PS & 2 & 1 & 0 & 1 & 0 & 0 & 0 & 0 & 0 \\
Percentage of All Transients &  & 50.00\% & 0 & 50.00\% & 0 & 0 & 0 & 0 & 0 \\
Percentage of Classified Transients &  &  & 0 & 100.00\% & 0 & 0 & 0 & 0 & 0
\enddata
\end{deluxetable*}

The full list of our candidates of interest can be found in Table \ref{tab:events} in Appendix C. The redshift distribution of all classified transients with a determined redshift on the TNS (244 events) is presented in Figure \ref{fig:redshifts}.\footnote{One classified transient, AT 2022amc, has no redshift determination, since its spectrum consists of a blue continuum with no clearly identifiable lines} While our queries can in principle find TDEs out to a redshift of $z\sim$0.16 \citep[using our magnitude limit of 19.5 and a TDE typical peak absolute magnitude of -20;][]{vanVelzen2020}, the median redshift of our classified candidates of interest is $z=0.069$, and all but one of the TDEs are at redshifts $z<0.04$ (the most distant TDE, AT\,2022csn at a redshift of $z=0.148$, is also more luminous than typical TDEs; Y. Dgany et al. in preparation). The reason that most classified events are much closer than our redshift limit is likely because nearby events are typically prioritized for spectroscopic classification over more distant events. At the median redshift, our angular cut of 0\farcs5 from the galaxy nucleus corresponds to a physical cut of $\sim$0.66 kpc \citep[assuming the cosmology of][]{Hinshaw2013}.

\begin{figure*}
\includegraphics[width=\textwidth]{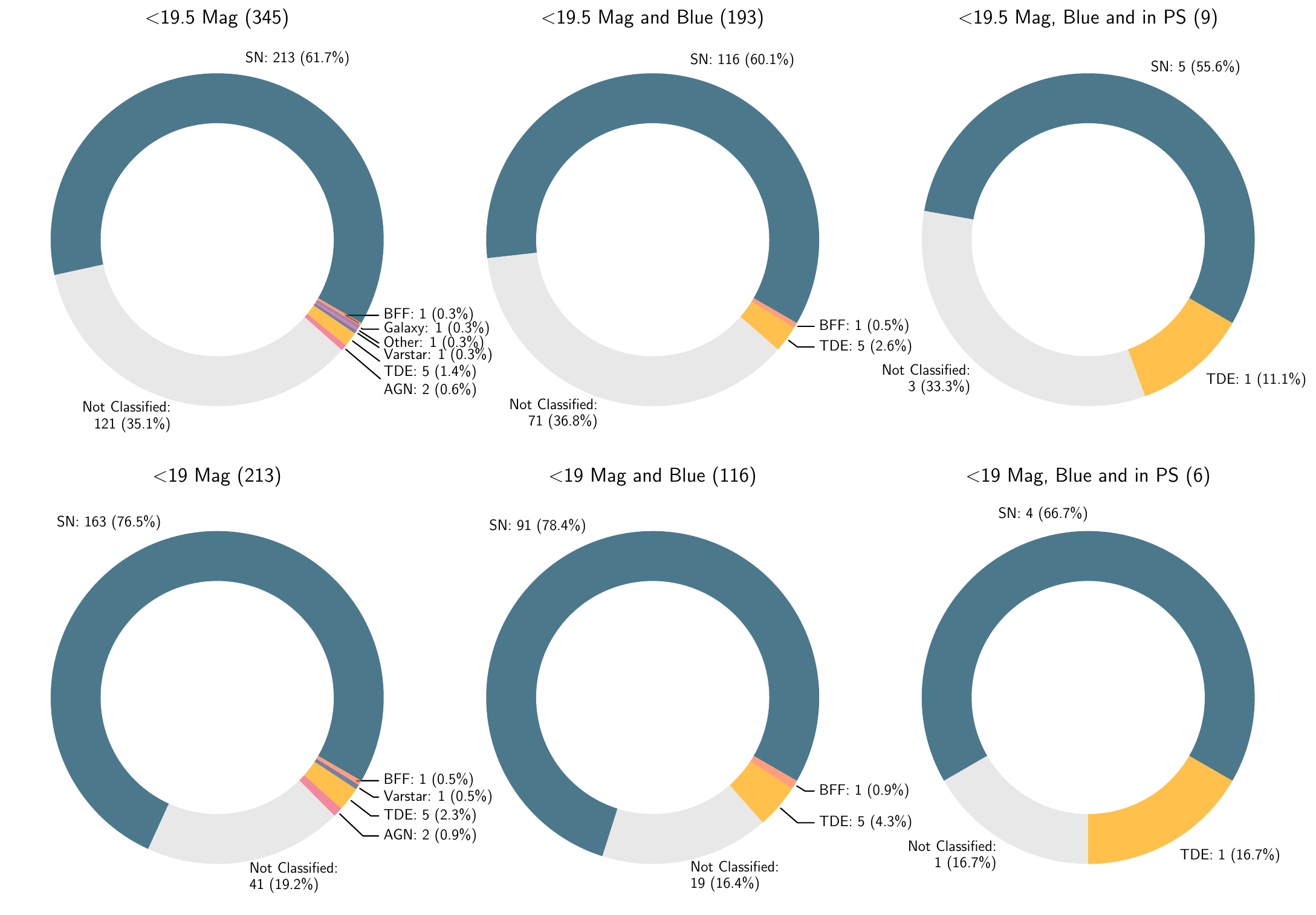}
\caption{\label{fig:pie}Numbers and fractions of the classes of candidates of interest from the different Lasair 1.0 queries, as detailed in the text and in Table \ref{tab:stats}. The numbers in parentheses next to each subplot title denote the total number of events in that query.}
\end{figure*}

\begin{figure*}[t]
\includegraphics[width=\textwidth]{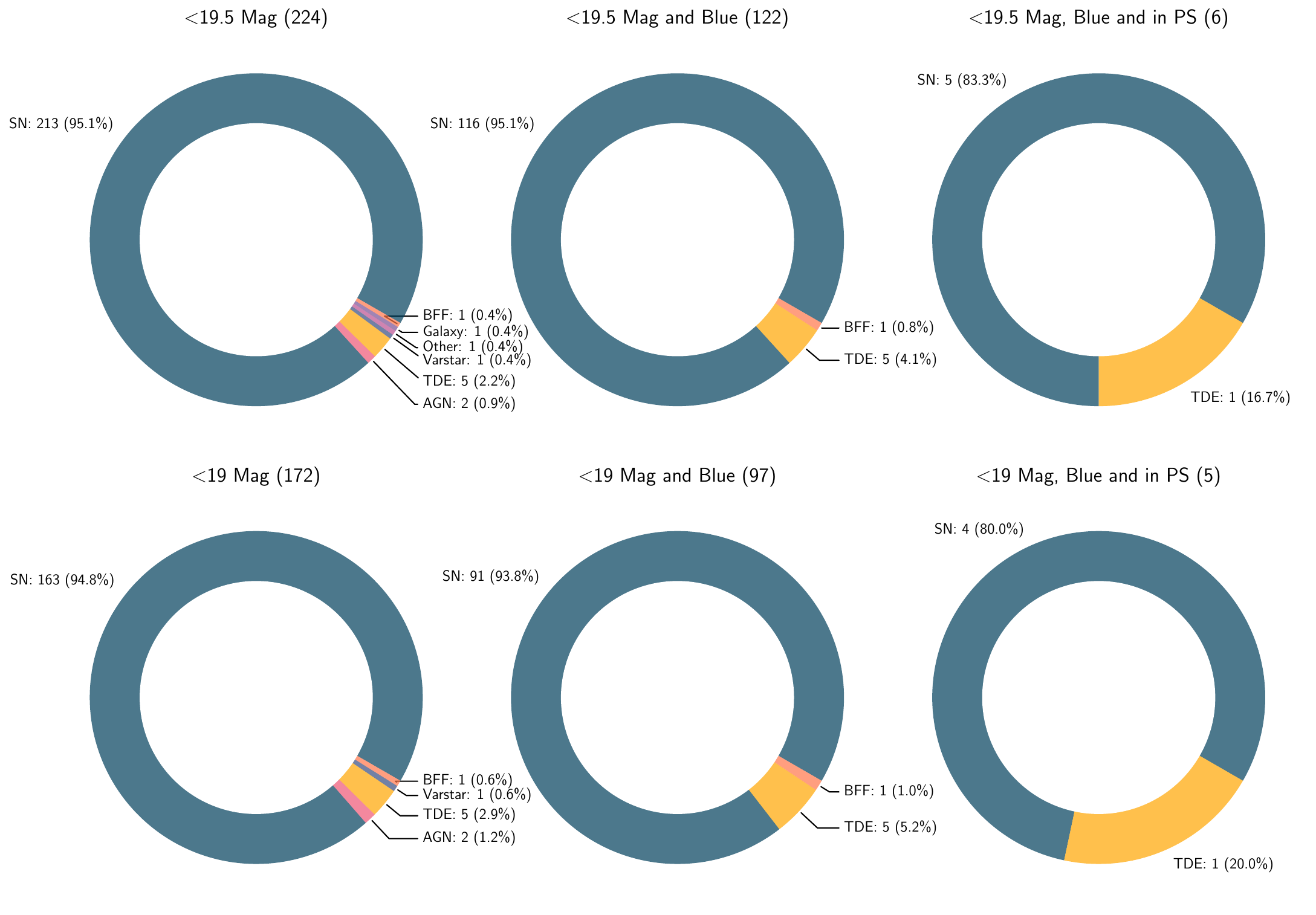}
\caption{\label{fig:pie_classified}The same as Figure \ref{fig:pie} but for classified events only.} 
\end{figure*}

The distribution of the classifications of our candidates of interest, per query, are listed in Table \ref{tab:stats} and presented in Figures \ref{fig:pie} and \ref{fig:pie_classified}. In the interest of simplicity, we consolidate the various SN classifications into one category, which we name ``SN''. These include SNe of undetermined type, SNe I of undetermined subtype, SNe Ia and their various subtypes, as well as SNe Ib, Ic, Ic-BL, II, IIn, IIb, and superluminous SNe (SLSNe) of Types I and II. A breakdown of the number of events per SN type is available in Table \ref{tab:snstats} in Appendix A.

For all of our events of interest in Lasair 1.0 (Query VI), we find that the vast majority (95.09\%) of classified events are SNe, 2.23\% (five events) are TDEs, 0.45\% are BFFs, and 0.89\% are flaring AGN.\footnote{Here, the term ``flaring AGN'' refers to events with coherent brightening episodes much stronger than any typical variability seen in their historical light curves. Specifically, the flares seen here rose by 0.25 magnitudes on average in one week, which is much higher than normal AGN variability \citep[e.g.][]{MacLeod2012,Caplar2017} } The remaining 1.34\% consist of one variable star, one event classified as ``Galaxy'' (which means that it was either an artifact or it faded before the spectrum was obtained), and one classified as ``Other''. The ``Other'' event is AT\,2022amc, which displays a featureless blue continuum. This could have been a young core-collapse SN or some other hot flare, including an SMBH-related one, such as a TDE or BFF. Unfortunately, no follow-up spectra were posted to TNS or, to our knowledge, published elsewhere, so its nature remains undetermined. 

The five TDEs are AT\,2020vwl \citep[ZTF20achpcvt, also named ATLAS20bdgk and Gaia20etp;][]{2021TNSCR.159....1H}, AT\,2021ehb \citep[ZTF21aanxhjv, also named ATLAS21jdy;][]{2021TNSAN.103....1G,2021TNSCR2295....1Y,2021TNSAN.309....1A,2022TNSCR2102....1Y}, AT\,2022bdw \citep[ZTF22aaahtqz, also named ATLAS22dth, Gaia22baj, and PS22avi;][]{2022TNSCR.511....1A,2022TNSAN..50....1A}, AT\,2022csn \citep[ZTF22aabimec, also named ATLAS22ggz, Gaia22ayp and PS22bju;][]{2022TNSCR3660....1A}, and AT\,2022dbl \citep[ZTF18aabdajx, also named ASASSN-22ci;][]{2022TNSCR.504....1A,2022TNSAN..57....1S}. AT\,2020vwl was classified as a ``TDE H+He'' \citep{vanVelzen2020} by the ZTF group, with a spectrum obtained from the Spectral Energy Distribution Machine \citep[SEDM;][]{Blagorodnova2018} on the Palomar 60 inch telescope \citep{Cenko2006}. Despite the low spectral resolution, broad He II and H$\alpha$ can be clearly identified, on top of a blue continuum, making the TDE classification secure. AT\,2021ehb was also classified by the ZTF team using an SEDM spectrum, but that spectrum is not publicly available on the TNS. Follow-up spectra that are available on the TNS do not show clear TDE signatures, but X-ray detections \citep{2021TNSAN.183....1Y} make the TDE classification likely. AT\,2022bdw, AT\,2022csn, and AT\,2022dbl were classified by the effort presented here, using the Las Cumbres Observatory Floyds spectrographs, based on broad H and He II spectral features on top of a blue continuum. AT\,2022csn was initially classified by the ePESSTO+ collaboration as a Type I SLSN \citep{2022TNSCR.667....1S,2022TNSAN..61....1S} but later reclassified by us as a TDE after the emergence of TDE spectral features. AT\,2022dbl is also listed on the TNS as AT\,2018mac, due to a sporadic detection in 2018 at a similar position, likely resulting from an image-subtraction artifact. This is the only TDE out of the five found in a PS host from the \cite{French2018} catalog. We conclude that all five TDE classifications are secure, with the possible exception of AT\,2021ehb, since its classification spectrum is not available on the TNS.

The BFF is AT\,2021seu \citep[ZTF21abjciua, also named ATLAS21bbfi;][]{2021TNSAN.195....1A}, also classified by this effort using Floyds \citep{2021TNSCR2460....1A}. The classification is based on a possible N III / He II emission complex on top of a blue continuum, not seen in an archival SDSS spectrum at that position \citep{2021TNSAN.195....1A}, and resembling the spectra of BFFs in \cite{Trakhtenbrot2019}. 

All five TDEs pass the ``blue criterion'' (Criterion 11) and were found by Query V, but the number of SNe passing this criterion is much smaller, nearly doubling the percentage of TDEs among classified blue transients. Limiting the search to candidates that are both blue and brighter than magnitude 19 at discovery (Query II) keeps all TDEs and further increases their percentage to 5.15\%. Looking at events in PS galaxies (Query IV),\footnote{Here, we study all events that were both in a PS host and blue. There were two more transients in PS hosts that were not blue: ZTF20acselme (a Type Ia SN) and ZTF22aabsemf (an unclassified event).} only one of the five TDEs remains, but it is one of six (16.67\%) classified transients there. This is consistent (to 1.2$\sigma$) with the finding of \cite{Arcavi2022} that $10.0\%\pm5.5\%$ of classified transients in PS galaxies should be TDEs.

For all of our events of interest in Lasair 3.0 (Query VIII), SNe are still the majority of classified events (72.73\%), with the rest all flaring AGNs.
The fraction of flaring AGNs in Lasair 3.0 is thus 30 times larger than in Lasair 1.0. Some of this difference is likely explained by the fact that, at least initially, Lasair 3.0 did not provide the full multiyear light-curve history of each candidate. This precluded filtering most AGNs by their historical activity.

In order to quantify the significance of the difference in fractions between queries, we calculate their confidence bounds using the Clopper-Pearson method \citep{clopper1934}. \cite{Gehrels1986} discusses how this method, which uses binomial statistics to estimate lower and upper confidence bounds for ratios, is especially useful for ratios of different event types, when the numbers of observed events are small. The $1\sigma$ confidence bounds calculated with this method (and used hereafter) are shown in Table \ref{tab:rates}. 

The fraction of TDEs in our global Lasair 1.0 query (Query VI) is $2.23\%\pm0.98\%$ and that of BFFs is $0.45\%\pm0.44$. Requiring candidates be blue (Query V), increases the TDE fraction by a factor of $1.84\pm1.11$. Adding the requirement for a PS host (Query IV) increases the TDE fraction by a factor of $7.47\pm7.53$ compared to the global query (Query VI). Without the full light-curve history of Lasair 3.0, the fraction of AGNs there increases by a factor of $30.55\pm23.86$ in Query VIII compared to Query VI. 

\begin{deluxetable*}{llllllll}
\tablecaption{\label{tab:rates}Fractions of classified candidates of interest from the different queries with $1\sigma$ Clopper-Pearson confidence bounds.}
\tablehead{\colhead{Query} & \colhead{SN} & \colhead{AGN} & \colhead{TDE} & \colhead{Other} & \colhead{Galaxy} & \colhead{BFF} & \colhead{Varstar}}
\startdata
I & $80.00\%\pm17.79\%$ & $0$ & $20.00\%\pm17.79\%$ & $0$ & $0$ & $0$ & $0$ \\
II & $93.81\%\pm2.43\%$ & $0$ & $5.15\%\pm2.23\%$ & $0$ & $1.03\%\pm1.02\%$ & $0$ & $0$ \\
III & $94.77\%\pm1.69\%$ & $1.16\%\pm0.81\%$ & $2.91\%\pm1.27\%$ & $0.58\%\pm0.58\%$ & $0.58\%\pm0.58\%$ & $0$ & $0$ \\
IV & $83.33\%\pm15.13\%$ & $0$ & $16.67\%\pm15.13\%$ & $0$ & $0$ & $0$ & $0$ \\
V & $95.08\%\pm1.95\%$ & $0$ & $4.10\%\pm1.78\%$ & $0$ & $0.82\%\pm0.81\%$ & $0$ & $0$ \\
VI & $95.09\%\pm1.44\%$ & $0.89\%\pm0.63\%$ & $2.23\%\pm0.98\%$ & $0.45\%\pm0.44\%$ & $0.45\%\pm0.44\%$ & $0.45\%\pm0.44\%$ & $0.45\%\pm0.44\%$ \\
VII & $70.00\%\pm10.19\%$ & $30.00\%\pm10.19\%$ & $0$ & $0$ & $0$ & $0$ & $0$ \\
VIII & $72.73\%\pm9.44\%$ & $27.27\%\pm9.44\%$ & $0$ & $0$ & $0$ & $0$ & $0$ \\
IX & $0$ & $100.00\%\pm0.00\%$ & $0$ & $0$ & $0$ & $0$ & $0$ \\
X & $0$ & $100.00\%\pm0.00\%$ & $0$ & $0$ & $0$ & $0$ & $0$
\enddata
\end{deluxetable*}

\begin{figure}
\centering
\includegraphics[width=\columnwidth]{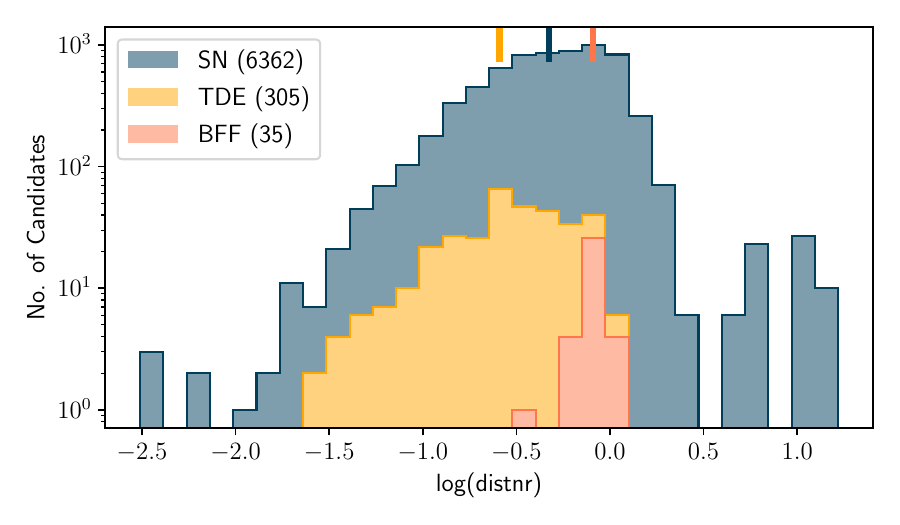}
\caption{\label{fig:distnr}Distribution of the {\tt distnr} parameter, which quantifies the offset of a source from its host galaxy center (shown in a stacked histogram) for all detections of SNe, TDEs, and the BFF. The vertical lines at the top denote the average value for each class. The number of total detections per class of events is shown in parentheses in the legend. While TDEs show a lower average offset compared to SNe, the difference is not large enough to be used a discriminant.} 
\end{figure}

We wish to check whether the offset of a source from its host galaxy center can be used as a way to better select for TDEs and BFFs. To do that, we retrieved the {\tt distnr} parameter of each detection of each SN, TDE, and the BFF in our sample using the Automatic Learning for the Rapid Classification of Events (ALeRCE) broker\footnote{\url{https://alerce.science/}} \citep{Forster2021}. The {\tt distnr} parameter is provided with each detection of a source in the ZTF alert packets\footnote{\url{https://zwickytransientfacility.github.io/ztf-avro-alert/schema.html}}. It denotes the distance of that detection to the nearest source in the reference-image PSF catalog (within $30\arcsec$), in units of pixels (which are equal to units of $\arcsec$, since the ZTF pixel scale is 1 pixel per $\arcsec$). We plot the distribution of these values in Figure \ref{fig:distnr}. TDEs show a slightly lower average {\tt distnr} value than SNe, but the difference is much smaller than the spread of values of each type of event and hence not significant. Our one BFF actually shows a larger average {\tt distnr} value than the SNe. However, this is based on detections of a single event, and could be driven by the centroid measurement of its particular host galaxy in ZTF. We conclude that there is no significant difference in the {\tt distnr} values between SNe, TDEs, and BFFs in ZTF, and therefore that this parameter is not a good discriminant. 

There are 24 additional events that are classified as TDEs on the TNS, from the time period of our search, which do not appear here. Of those, 16 have robust TDE classifications (i.e. at least one public spectrum showing clear broad He II and/or H$\alpha$ emission and blue colors). The rest have either no public spectrum available, very noisy spectra, or show no clear spectral features. Of the 16 robust TDEs, most (11) have not been identified here because they were deemed not to have a coherent rising light curve at discovery. Three events were missed due to a bug in the queries (which was later fixed), and two did not pass Criterion 7, regarding the number of unreliable detections (one required increasing the number of unreliable detections from $<$3 to $<$5, and one required it be increased to $<$21). However, relaxing Criterion 7 would have likely also led to an increase in contaminant candidates.

\section{Discussion and Conclusions}\label{sec:conclusions}

Our results quantify the ``needle in a haystack'' problem of finding TDEs and BFFs in wide-field transient surveys. We find that the photometric history of candidates is crucial for removing most AGN contamination. Even so, roughly one in 35--45 events is a TDE, and one in 170--220 is a BFF. This sets a significant challenge for identifying these events in current transient surveys, and for identifying even a small subset of the thousands of TDEs expected to be discovered by the Legacy Survey of Space and Time \citep[LSST;][]{LSST} every year \citep{Bricman2020}.   

The fraction of TDEs increases by almost a factor of 2 to roughly one in 20--25 events when selecting only blue ($g-r<0.05$) transients. This cut does not remove any TDEs. The fraction further increases by another factor of $\sim$3 to roughly one in 5--6 events when selecting probable TDE host galaxies. However, such galaxies are rare, making the total number of TDEs discoverable in this way small. Given the huge increase in expected TDE discovery fractions, though, it would be beneficial to update the \cite{French2018} galaxy catalog and to expand its coverage using new spectroscopic and photometric surveys such as SDSS-V \citep{Kollmeier2017}. 

An additional $\sim$50\% of TDEs are found when relaxing Criterion 7 to allow events with a smaller number of reliable detections, and roughly three times as many TDEs are found when relaxing the condition that the event be rising in brightness at discovery. However, the number of contaminants that relaxing such criteria adds is significant. For example, changing the threshold of Criterion 7 from $<$3 to $<$5 (which would have added one TDE to the sample) increased the number of daily candidates by a factor of 2--3, and changing it to $<$21 (which would have added a second TDE to the sample) increased the number of daily candidates by an order of magnitude.  

We find no significant difference between the offsets of TDEs vs. nuclear SNe from their host galaxy centers. Therefore, this parameter cannot be used as a discriminant for selecting more likely TDEs, at least not in ZTF, as quantified by the {\tt distnr} parameter. LSST, with its higher spatial resolution, might be able to make host nucleus offsets a more viable distinguishing parameter. 

An additional possible discriminant for selecting TDEs, but which was not tested here, is their ultraviolet to optical colors \citep[e.g.][]{Hung2018}. Obtaining ultraviolet photometry rapidly and for many targets is currently possible almost exclusively with the the Neil Gehrels {\it Swift} Observatory \citep{Gehrels2004} Ultraviolet/Optical Telescope \citep{Roming2005}. Indeed, \cite{Hung2018} have shown that selecting transients by a combination of their ultraviolet to optical colors using {\it Swift} and the optical colors of their host galaxies increased the fraction of TDEs to 1 in 4.5. However, {\it Swift} is limited in the number of transients it can vet. The upcoming Ultraviolet Transient Astronomy Satellite \citep[ULTRASAT;][]{Sagiv2014}, with its wide-field ultraviolet imager, is expected to obtain ultraviolet photometry for thousands of TDEs. This will be an excellent way to discriminate TDEs from other transients without the need for substantial classification resources.

Another approach is to train machine-learning algorithms to classify transients from photometry alone. This has been done for distinguishing between some SN types \citep[e.g.][]{Richards2012,Moller2016,Charnock2017,Boone2019,Ishida2019,Villar2019,Gomez2020,Hosseinzadeh2020,Villar2020}. Until recently, the number of observed TDEs has been too small to be used for classification training, leading \cite{Muthukrishna2019} to train their algorithm on simulated TDE light curves. Today, with TDE light curves available for dozens of events, it might be possible to effectively train a machine-learning algorithm to distinguish TDE light curves from those of other SNe. How effectively this can be done, and at what phase of the light curve a robust classification can be obtained (if at all), remains to be tested.

Of course, any such filtering based on the photometric properties of the transient or the characteristics of its host galaxy can also bias population studies of TDEs and BFFs. A specific population that almost all current searches (including ours) are biased against is that of slowly evolving transients with years-long evolution. Such events are already suppressed by the alert mechanisms of most transient surveys, even before reaching the brokers. ZTF alerts, for example, are generated by comparing a new image to a reference image taken up to a few months to a few years earlier. Therefore, events that rise on a time scale of several years will not be much brighter in the new image compared to the reference, and thus an alert might never be issued. Since the set of images used as references is updated from time to time, such transients could remain hidden during the lifetimes of the surveys. Indeed, when \cite{Lawrence2016} compared images from PS1 to images obtained a decade earlier by SDSS, they found a population of slowly rising nuclear transients. This population is not seen in current transient surveys, which are optimized to find transients that change on shorter time scales.

It will thus continue to be challenging to find TDEs and BFFs in optical transient surveys in an unbiased way, even for events evolving on time scales of days to months. One way forward is to use some combination of well-defined photometric and host galaxy filters, such as those used here. However, making searches as complete as possible will still require ample spectroscopic resources for vetting large numbers of SMBH-related transient candidates. 

~\\
We thank B. Trakhtenbrot for providing some of the spectroscopy time on the Las Cumbres network used to classify candidates identified here, and A. Lawrence, K. Smith, R. Williams, and D. Young for assistance with using Lasair and for helpful comments. We also thank M. Nicholl for implementing many of the queries used here on Lasair, as well as the \cite{French2018} galaxy catalog as a watchlist there, and A. Riba for developing a Target and Observation Manager used to manually inspect candidates and schedule followup observations. We are grateful to B. Zackay for helpful comments regarding host offsets. 

Y.D., I.A., and L.M. acknowledge support from the European Research Council (ERC) under the European Union's Horizon 2020 research and innovation program (grant agreement number 852097).
I.A. is a CIFAR Azrieli Global Scholar in the Gravity and the Extreme Universe Program and acknowledges support from that program, from the Israel Science Foundation (grant No. 2752/19), from the United States -- Israel Binational Science Foundation (BSF; grant No. 2018166), and from the Israeli Council for Higher Education Alon Fellowship.
This work makes use of observations with the Las Cumbres Observatory global telescope network. The Las Cumbres Observatory group is supported by NSF grants AST-1911151 and AST-1911225 and BSF grant 2018166.

\bibliography{refs}{}
\bibliographystyle{aasjournal}

\appendix

\section{Fractions of SN Subtypes}\label{sec:sne}

We list the division of SN types found by each query in Table \ref{tab:snstats}. The majority of contaminants across all queries are Type Ia SNe, with the next most likely contaminant being Type II SNe (including their subvariants IIb and IIn). The ZTF Bright Transient Survey \citep[BTS;][]{Perley2020} aims to classify all transients brighter than certain magnitude cuts (similar to our Query III), but, unlike this work, does not focus on galaxy nuclei. Their Type~Ia SN fraction is lower than ours ($\sim$73\% vs. $\sim$82\% here) and their Type~II SN fraction is higher ($\sim$20\% vs. $\sim$10\% here)\footnote{\url{https://sites.astro.caltech.edu/ztf/bts/bts.php}}. Since Type~Ia and Type~II SNe have similar radial distributions in their host galaxies \citep[e.g.][]{Prieto2008}, this difference might be a selection effect whereby Type Ia SNe are preferentially detected in this work compared to the BTS, since they are more luminous than Type II events and stand out more clearly in bright galaxy nuclei.
As found previously by \cite{Arcavi2022}, Type Ia SNe are the only contaminant of TDEs and BFFs in PS galaxies (unless AGNs cannot be fully filtered out, as in the case of the Lasair 3.0 queries).  

\begin{deluxetable}{llllllllllll}[h]
\tablecaption{\label{tab:snstats}Internal division of SN types from the different queries.}
\tablehead{
\colhead{Query} & \colhead{Total} & \colhead{SN Ia} & \colhead{SN Ib} & \colhead{SN Ic} & \colhead{SN Ic-BL} & \colhead{SN I} & \colhead{SN II} & \colhead{SN IIb} & \colhead{SN IIn} & \colhead{SN} & \colhead{SLSN}\\
\colhead{} & \colhead{SNe} & \colhead{} & \colhead{} & \colhead{} & \colhead{} & \colhead{} & \colhead{} & \colhead{} & \colhead{} & \colhead{} & \colhead{}}
\startdata
I & 4 & 4 (100.0\%) & 0 & 0 & 0 & 0 & 0 & 0 & 0 & 0 & 0 \\
II & 91 & 76 (83.5\%) & 1 (1.1\%) & 0 & 2 (2.2\%) & 0 & 10 (11.0\%) & 0 & 2 (2.2\%) & 0 & 0 \\
III & 163 & 134 (82.2\%) & 2 (1.2\%) & 2 (1.2\%) & 4 (2.5\%) & 0 & 17 (10.4\%) & 0 & 3 (1.8\%) & 1 (0.6\%) & 0 \\
IV & 5 & 5 (100.0\%) & 0 & 0 & 0 & 0 & 0 & 0 & 0 & 0 & 0 \\
V & 116 & 97 (83.6\%) & 1 (0.9\%) & 0 & 2 (1.7\%) & 0 & 13 (11.2\%) & 0 & 2 (1.7\%) & 0 & 1 (0.9\%) \\
VI & 213 & 175 (82.2\%) & 2 (0.9\%) & 3 (1.4\%) & 4 (1.9\%) & 1 (0.5\%) & 21 (9.9\%) & 1 (0.5\%) & 3 (1.4\%) & 1 (0.5\%) & 2 (0.9\%) \\
VII & 14 & 8 (57.1\%) & 0 & 0 & 0 & 0 & 4 (28.6\%) & 0 & 2 (14.3\%) & 0 & 0 \\
VIII & 16 & 10 (62.5\%) & 0 & 0 & 0 & 0 & 4 (25.0\%) & 0 & 2 (12.5\%) & 0 & 0 \\
IX & 0 & 0 & 0 & 0 & 0 & 0 & 0 & 0 & 0 & 0 & 0 \\
X & 0 & 0 & 0 & 0 & 0 & 0 & 0 & 0 & 0 & 0 & 0
\enddata
\end{deluxetable}

\section{Results per Separation Threshold}\label{sec:sne}

As mentioned in Section \ref{sec:methods}, we split the Lasair 1.0 sample into two subsamples, each with a different separation threshold for Condition 1 (0\farcs5 and 1\arcsec). The fractions (Table \ref{tab:stats}) of the split per separation threshold are presented in Table \ref{tab:stats0.5} (for the 0\farcs5 threshold) and Table \ref{tab:stats1} (for the 1\arcsec\ threshold). 

\begin{deluxetable}{llllllllll}
\tablecaption{\label{tab:stats0.5}The same as Table \ref{tab:stats} but only for events in Lasair 1.0 selected with a separation (Condition 1) threshold of 0\farcs5.}
\tablehead{
\colhead{Query} & \colhead{Total} & \colhead{Not} & \colhead{SN} & \colhead{AGN} & \colhead{TDE} & \colhead{Other} & \colhead{Galaxy} & \colhead{BFF} & \colhead{Varstar}\\
\colhead{} & \colhead{Transients} & \colhead{Classified} & \colhead{} & \colhead{} & \colhead{} & \colhead{} & \colhead{} & \colhead{} & \colhead{}}
\startdata
\multicolumn{10}{c}{Lasair 1.0} \\
\hline
I: $<$19 Mag, Blue, and in PS & 4 & 1 & 3 & 0 & 0 & 0 & 0 & 0 & 0 \\
Percentage of All Transients &  & 25.00\% & 75.00\% & 0 & 0 & 0 & 0 & 0 & 0 \\
Percentage of Classified Transients &  &  & 100.00\% & 0 & 0 & 0 & 0 & 0 & 0 \\
\hline
II: $<$19 Mag and Blue & 96 & 17 & 76 & 0 & 2 & 0 & 1 & 0 & 0 \\
Percentage of All Transients &  & 17.71\% & 79.17\% & 0 & 2.08\% & 0 & 1.04\% & 0 & 0 \\
Percentage of Classified Transients &  &  & 96.20\% & 0 & 2.53\% & 0 & 1.27\% & 0 & 0 \\
\hline
III: $<$19 Mag & 186 & 38 & 143 & 1 & 2 & 1 & 1 & 0 & 0 \\
Percentage of All Transients &  & 20.43\% & 76.88\% & 0.54\% & 1.08\% & 0.54\% & 0.54\% & 0 & 0 \\
Percentage of Classified Transients &  &  & 96.62\% & 0.68\% & 1.35\% & 0.68\% & 0.68\% & 0 & 0 \\
\hline
IV: $<$19.5 Mag, Blue, and in PS & 7 & 3 & 4 & 0 & 0 & 0 & 0 & 0 & 0 \\
Percentage of All Transients &  & 42.86\% & 57.14\% & 0 & 0 & 0 & 0 & 0 & 0 \\
Percentage of Classified Transients &  &  & 100.00\% & 0 & 0 & 0 & 0 & 0 & 0 \\
\hline
V: $<$19.5 Mag and Blue & 169 & 67 & 99 & 0 & 2 & 0 & 1 & 0 & 0 \\
Percentage of All Transients &  & 39.64\% & 58.58\% & 0 & 1.18\% & 0 & 0.59\% & 0 & 0 \\
Percentage of Classified Transients &  &  & 97.06\% & 0 & 1.96\% & 0 & 0.98\% & 0 & 0 \\
\hline
VI: $<$19.5 Mag & 310 & 113 & 191 & 1 & 2 & 1 & 1 & 1 & 0 \\
Percentage of All Transients &  & 36.45\% & 61.61\% & 0.32\% & 0.65\% & 0.32\% & 0.32\% & 0.32\% & 0 \\
Percentage of Classified Transients &  &  & 96.95\% & 0.51\% & 1.02\% & 0.51\% & 0.51\% & 0.51\% & 0 \\
\enddata
\end{deluxetable}

\begin{deluxetable}{llllllllll}
\tablecaption{\label{tab:stats1}The same as Table \ref{tab:stats} but only for events in Lasair 1.0 selected with a separation (Condition 1) threshold of 1\arcsec.}
\tablehead{
\colhead{Query} & \colhead{Total} & \colhead{Not} & \colhead{SN} & \colhead{AGN} & \colhead{TDE} & \colhead{Other} & \colhead{Galaxy} & \colhead{BFF} & \colhead{Varstar}\\
\colhead{} & \colhead{Transients} & \colhead{Classified} & \colhead{} & \colhead{} & \colhead{} & \colhead{} & \colhead{} & \colhead{} & \colhead{}}
\startdata
\multicolumn{10}{c}{Lasair 1.0} \\
\hline
I: $<$19 Mag, Blue, and in PS & 2 & 0 & 1 & 0 & 1 & 0 & 0 & 0 & 0 \\
Percentage of All Transients &  & 0 & 50.00\% & 0 & 50.00\% & 0 & 0 & 0 & 0 \\
Percentage of Classified Transients &  &  & 50.00\% & 0 & 50.00\% & 0 & 0 & 0 & 0 \\
\hline
II: $<$19 Mag and Blue & 20 & 2 & 15 & 0 & 3 & 0 & 0 & 0 & 0 \\
Percentage of All Transients &  & 10.00\% & 75.00\% & 0 & 15.00\% & 0 & 0 & 0 & 0 \\
Percentage of Classified Transients &  &  & 83.33\% & 0 & 16.67\% & 0 & 0 & 0 & 0 \\
\hline
III: $<$19 Mag & 27 & 3 & 20 & 1 & 3 & 0 & 0 & 0 & 0 \\
Percentage of All Transients &  & 11.11\% & 74.07\% & 3.70\% & 11.11\% & 0 & 0 & 0 & 0 \\
Percentage of Classified Transients &  &  & 83.33\% & 4.17\% & 12.50\% & 0 & 0 & 0 & 0 \\
\hline
IV: $<$19.5 Mag, Blue, and in PS & 2 & 0 & 1 & 0 & 1 & 0 & 0 & 0 & 0 \\
Percentage of All Transients &  & 0 & 50.00\% & 0 & 50.00\% & 0 & 0 & 0 & 0 \\
Percentage of Classified Transients &  &  & 50.00\% & 0 & 50.00\% & 0 & 0 & 0 & 0 \\
\hline
V: $<$19.5 Mag and Blue & 24 & 4 & 17 & 0 & 3 & 0 & 0 & 0 & 0 \\
Percentage of All Transients &  & 16.67\% & 70.83\% & 0 & 12.50\% & 0 & 0 & 0 & 0 \\
Percentage of Classified Transients &  &  & 85.00\% & 0 & 15.00\% & 0 & 0 & 0 & 0 \\
\hline
VI: $<$19.5 Mag & 35 & 8 & 22 & 1 & 3 & 0 & 0 & 0 & 1 \\
Percentage of All Transients &  & 22.86\% & 62.86\% & 2.86\% & 8.57\% & 0 & 0 & 0 & 2.86\% \\
Percentage of Classified Transients &  &  & 81.48\% & 3.70\% & 11.11\% & 0 & 0 & 0 & 3.70\% \\
\enddata
\end{deluxetable}

In order to assess if there is a statistically significant difference between the fractions resulting from the different thresholds, we calculate the 1$\sigma$ Clopper-Pearson confidence bounds per each separation threshold subsample in Table \ref{tab:rates0.5} (for the 0\farcs5 threshold) and Table \ref{tab:rates1} (for the 1\arcsec\ threshold).

\begin{deluxetable}{llllllll}
\tablecaption{\label{tab:rates0.5}The same as Table \ref{tab:rates} but only for events in Lasair 1.0 selected with a separation (Condition 1) threshold of 0\farcs5.}
\tablehead{\colhead{Query} & \colhead{SN} & \colhead{AGN} & \colhead{TDE} & \colhead{Other} & \colhead{Galaxy} & \colhead{BFF} & \colhead{Varstar}}
\startdata
I & $100.00\%\pm0.00\%$ & $0$ & $0$ & $0$ & $0$ & $0$ & $0$ \\
II & $96.20\%\pm2.14\%$ & $0$ & $2.53\%\pm1.76\%$ & $0$ & $1.27\%\pm1.25\%$ & $0$ & $0$ \\
III & $96.62\%\pm1.48\%$ & $0.68\%\pm0.67\%$ & $1.35\%\pm0.94\%$ & $0.68\%\pm0.67\%$ & $0.68\%\pm0.67\%$ & $0$ & $0$ \\
IV & $100.00\%\pm0.00\%$ & $0$ & $0$ & $0$ & $0$ & $0$ & $0$ \\
V & $97.06\%\pm1.66\%$ & $0$ & $1.96\%\pm1.37\%$ & $0$ & $0.98\%\pm0.97\%$ & $0$ & $0$ \\
VI & $96.95\%\pm1.22\%$ & $0.51\%\pm0.50\%$ & $1.02\%\pm0.71\%$ & $0.51\%\pm0.50\%$ & $0.51\%\pm0.50\%$ & $0.51\%\pm0.50\%$ & $0$ \\
\enddata
\end{deluxetable}

\begin{deluxetable}{llllllll}
\tablecaption{\label{tab:rates1}The same as Table \ref{tab:rates} but only for events in Lasair 1.0 selected with a separation (Condition 1) threshold of 1\arcsec.}
\tablehead{\colhead{Query} & \colhead{SN} & \colhead{AGN} & \colhead{TDE} & \colhead{Other} & \colhead{Galaxy} & \colhead{BFF} & \colhead{Varstar}}
\startdata
I & $50.00\%\pm35.16\%$ & $0$ & $50.00\%\pm35.16\%$ & $0$ & $0$ & $0$ & $0$ \\
II & $83.33\%\pm8.74\%$ & $0$ & $16.67\%\pm8.74\%$ & $0$ & $0$ & $0$ & $0$ \\
III & $83.33\%\pm7.57\%$ & $4.17\%\pm4.06\%$ & $12.50\%\pm6.71\%$ & $0$ & $0$ & $0$ & $0$ \\
IV & $50.00\%\pm35.16\%$ & $0$ & $50.00\%\pm35.16\%$ & $0$ & $0$ & $0$ & $0$ \\
V & $85.00\%\pm7.94\%$ & $0$ & $15.00\%\pm7.94\%$ & $0$ & $0$ & $0$ & $0$ \\
VI & $81.48\%\pm7.43\%$ & $3.70\%\pm3.61\%$ & $11.11\%\pm6.01\%$ & $0$ & $0$ & $0$ & $3.70\%\pm3.61\%$ \\
\enddata
\end{deluxetable}

\begin{figure}
\centering
\includegraphics[width=0.5\columnwidth]{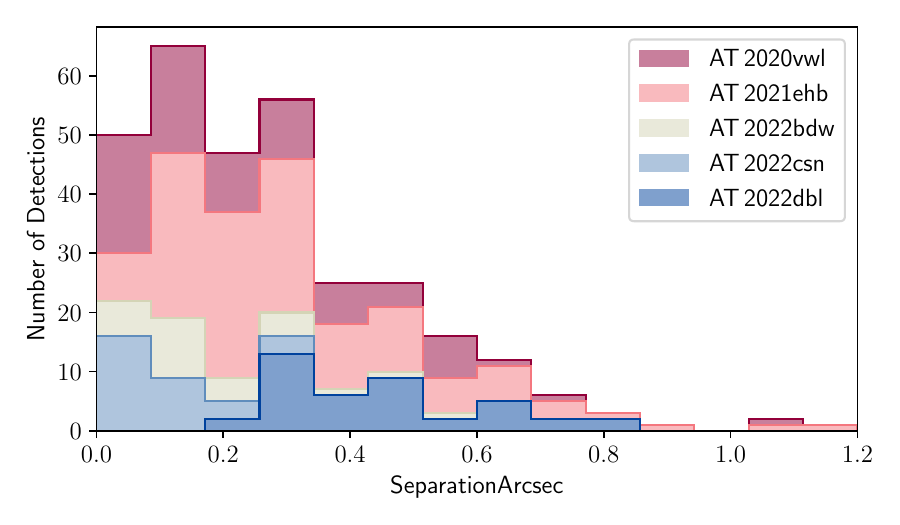}
\caption{\label{fig:separcsec}Distribution of the {\tt separationArcsec} parameter from Criterion 1 (shown in a stacked histogram) for all detections of our five TDEs. There is no significant difference in the separations between the TDEs in the 0\farcs5 threshold subsample (AT\,2020vwl and AT\,2021ehb) and the TDEs in the 1\arcsec\ threshold subsample (AT\,2022bdw, AT\,2022csn, and AT\,2022dbl), meaning that the change of separation threshold does not have a strong effect on the number of TDEs discovered.} 
\end{figure}

There are no statistically significant differences between the results of the two separation thresholds. It does appear that there is a much higher fraction of TDEs in the subsample of the 1\arcsec\ threshold (of order 10\%) compared to the 0\farcs5 threshold (of order 1\%). However, this is not statistically significant and is a result of small number statistics. To demonstrate this, we plot all of the \texttt{separationArcsec} values of all the detections of our five TDEs in Figure \ref{fig:separcsec}.

Most detections for the two TDEs in the 0\farcs5 threshold subsample and the three TDEs in the 1\arcsec\ threshold subsample are below 0\farcs5 separation. This means that the increase of the threshold to 1\arcsec\ is not responsible for the larger fraction of TDEs in that subsample, but rather it is a small number statistics fluctuation. 

\section{List of Events}\label{sec:all_events}

We list in Table \ref{tab:events} the full set of events considered transients of interest in this work, their publicly available classification and redshift, and the query number(s) in which they were found. 

\begin{longrotatetable}
\begin{deluxetable}{lp{0.09\linewidth}p{0.1\linewidth}DDllllp{0.22\linewidth}}
\centerwidetable
\tabletypesize{\footnotesize}
\tablecaption{\label{tab:events}Candidates of interest.}

\tablehead{
\colhead{ZTF Name} & \colhead{TNS Name} & \colhead{Other Names} & \multicolumn2c{R.A.} & \multicolumn2c{Decl.} & \colhead{Query} & \colhead{Classification} & \colhead{Redshift} & \colhead{Note} & \colhead{Reference(s)}\\
\colhead{} & \colhead{} & \colhead{} & \multicolumn2c{(deg)} & \multicolumn2c{(deg)} & \colhead{} & \colhead{} & \colhead{} & \colhead{} & \colhead{}}

\decimals
\startdata
ZTF18aabdajx & AT\,2022dbl (also AT\,2018mac) & ASASSN-22ci & 185.1878 & 49.55128 & I & TDE & 0.0284 &  & \cite{2022TNSCR.504....1A,2022TNSAN..57....1S} \\
ZTF18aacnlxz & SN\,2020aavr &  & 134.95468 & 38.10909 & III & SN II & 0.072475 &  & \cite{2020TNSCR3662....1B} \\
ZTF18aadsuxd & AT\,2020yui & ATLAS20bfcp, PS20kyb, ZTF18aaefvaq & 129.53396 & 31.66792 & III &  &  &  &  \\
ZTF18aagtwyh & SN\,2021oud & Gaia21cyq, ZTF18aahbypm & 189.92466 & 16.53793 & II & SN Ia & 0.066041 &  & \cite{2021TNSCR2021....1S} \\
ZTF18aahfbqp & SN\,2020acua &  & 156.67845 & 21.7239 & III & SN Ia & 0.041362 &  & \cite{2020TNSCR3921....1H} \\
ZTF18aahkoqq & SN\,2021slw & ATLAS21bbji, Gaia21dnq, ZTF21abnabtv & 203.17888 & 32.4431 & III & SN Ia & 0.034457 &  & \cite{2021TNSCR2457....1H,2021TNSCR3263....1C} \\
ZTF18aahvpcy & SN\,2022eaz & ATLAS22ina, Gaia22bhf & 208.40005 & 37.73573 & II & SN Ia & 0.0607 &  & \cite{2022TNSCR.843....1G} \\
ZTF18aaisqmw & SN\,2020acqt &  & 176.35752 & 55.52702 & III & SN Ia & 0.0527 &  & \cite{2020TNSCR3963....1B} \\
ZTF18aaiwewk & SN\,2022cox & ATLAS22gie, PS22biu & 196.64354 & 54.9627 & V & SN Ia & 0.087 &  & \cite{2022TNSCR.501....1P} \\
ZTF18aaizgoq & SN\,2021gad & ATLAS21ipu & 231.51665 & 36.32309 & III & SN Ia & 0.07 &  & \cite{2021TNSCR.913....1T} \\
ZTF18aajhkjz & AT\,2020acub & ATLAS20bkaw, PS20nox, ZTF20acxtayx & 149.88365 & 15.84735 & V &  &  &  &  \\
ZTF18aajhkjz & AT\,2021mrp &  & 166.45372 & 15.12587 & V &  &  &  &  \\
ZTF18aakeanj & AT\,2021dc & Gaia21ald, ZTF21aaaollj & 203.76788 & 53.87351 & II &  &  & \tablenotemark{a} &  \\
ZTF18aamtiwb & SN\,2022dcf & ATLAS22gxr, PS22cdu & 185.06648 & 56.62635 & II & SN Ia-91T-like & 0.066 &  & \cite{2022TNSCR.542....1P} \\
ZTF18aansqom & SN\,2021hbg &  & 251.0451 & 39.78875 & III & SN Ia & 0.03 &  & \cite{2021TNSCR1179....1A} \\
ZTF18aarefcm & AT\,2022drv & ATLAS22iga & 204.58714 & 40.15554 & II &  &  &  &  \\
ZTF18aasypqx & SN\,2022cvt & ATLAS22gxa & 214.68999 & 22.36559 & II & SN Ia-91T-like & 0.073 &  & \cite{2022TNSCR.571....1S,2022TNSCR.583....1S,2022TNSAN..54....1S} \\
ZTF18aauvmnq & SN\,2021tty & ATLAS21bden & 207.19808 & 26.55212 & II & SN Ia & 0.0621 &  & \cite{2021TNSCR2575....1G,2021TNSCR2623....1C} \\
ZTF18aaviokz & AT\,2021duz &  & 207.86824 & 40.44794 & IV &  &  &  &  \\
ZTF18aaviokz & AT\,2021duz &  & 207.86828 & 40.44793 & I &  &  & \tablenotemark{c} &  \\
ZTF18abfdxbt & SN\,2022dme & Gaia22bak & 255.92866 & 78.65643 & II & SN Ia & 0.048914 &  & \cite{2022TNSCR.719....1B} \\
ZTF18abmakin &  &  & 298.62199 & 55.58944 & II &  &  &  &  \\
ZTF18abrfccm & SN\,2020yxo & ATLAS20beyo & 26.86302 & 0.96765 & III & SN Ia & 0.1 &  & \cite{2020TNSCR3382....1G,2020TNSAN.222....1C} \\
ZTF18abrzevn & AT\,2019byp &  & 269.35527 & 26.84119 & VII & AGN & 0.045399 &  & \cite{2022TNSCR1275....1N} \\
ZTF18abwitkf & SN\,2021xy & ATLAS21bsl, Gaia21amh & 15.70534 & -21.83647 & III & SN Ia & 0.061 &  & \cite{2021TNSCR.157....1S,2021TNSAN..18....1S} \\
ZTF18abwnmnc & AT\,2020aczj & ZTF20acxuyqs & 330.52799 & -5.66265 & VII &  &  &  &  \\
ZTF18abxusti & AT\,2021gvi &  & 119.03849 & 65.87328 & III &  &  &  &  \\
ZTF18acfhmhp & AT\,2020aaxb &  & 13.13143 & 46.33825 & III &  &  &  &  \\
ZTF18acjxyak & SN\,2021efe &  & 281.14242 & 33.72181 & III & SN Ia & 0.0585 &  & \cite{2021TNSCR.849....1P} \\
ZTF18acsremz & SN\,2021hmc &  & 148.62747 & 53.16909 & I & SN Ia & 0.025136 &  & \cite{2021TNSCR1047....1B,2021TNSCR1102....1D} \\
ZTF18actytgu & SN\,2021lyz & ATLAS21barz & 11.03286 & 30.351 & II & SN Ia & 0.019 &  & \cite{2021TNSCR1750....1T} \\
ZTF18acwyvak & AT\,2021jmb &  & 174.04786 & 74.77397 & VII & AGN & 0.055785 &  & \cite{2022TNSCR1275....1N} \\
ZTF20aanlygx & AT\,2020abea &  & 56.01208 & -18.62466 & III & AGN & 0.063 &  & \cite{2023TNSCR.853....1A} \\
ZTF20aanzldk & AT\,2022cwl & ATLAS22gfj, PS22bps & 139.37062 & -2.08112 & III & AGN & 0.053 &  & \cite{2022TNSCR.512....1A} \\
ZTF20aaoluli & AT\,2021aqe &  & 14.09962 & 13.90157 & VI &  &  &  &  \\
ZTF20aaprrgg & SN\,2021aayn & PS21oli & 62.01957 & -6.30023 & IV & SN Ia & 0.087 &  & \cite{2021TNSCR3519....1C,2021TNSAN.265....1C} \\
ZTF20aaqhlvq & AT\,2022jiw & ATLAS22nmb, Gaia22dmk & 196.76539 & -4.30944 & VII &  &  &  &  \\
ZTF20aaqhlvq & AT\,2022jiw & ATLAS22nmb, Gaia22dmk & 196.7654 & -4.30943 & VIII &  &  &  &  \\
ZTF20aartqux & AT\,2021hed & ATLAS21jyx & 208.60376 & 23.67644 & II &  &  &  &  \\
ZTF20aarvpbq & SN\,2021gen & ATLAS21ikj & 222.98006 & 10.12371 & III & SN Ia & 0.080611 &  & \cite{2021TNSCR.922....1D} \\
ZTF20aasivtu & AT\,2020fhp & Gaia20dxw & 276.63611 & 21.15216 & VII & AGN & 0.223 &  & \cite{2022TNSCR1320....1N} \\
ZTF20aavfogu & AT\,2020hoj &  & 173.45788 & -20.81444 & VI &  &  & \tablenotemark{a} &  \\
ZTF20abajndx & AT\,2020acvj &  & 187.25093 & -15.50851 & II &  &  & \tablenotemark{a} &  \\
ZTF20abbhlct & AT\,2020mma &  & 188.91823 & -21.04658 & VII & AGN & 0.061759 &  & \cite{2022TNSCR1275....1N} \\
ZTF20abeyroj & AT\,2022ebh & Gaia22dfz & 198.5069 & -3.47522 & VII &  &  &  &  \\
ZTF20abmtbpa & AT\,2021adlq & ATLAS21blqh & 31.75077 & 6.96475 & VI &  &  &  &  \\
ZTF20abpbmxi & AT\,2020qnk &  & 297.46295 & -18.15618 & II &  &  & \tablenotemark{b} &  \\
ZTF20absulsv & AT\,2021adhi & ATLAS21blnl, PS21mqk & 358.86024 & -4.2969 & V &  &  &  &  \\
ZTF20abtcijp & SN\,2021iat & ATLAS21kmm & 258.99041 & 66.6909 & II & SN Ia & 0.07 &  & \cite{2021TNSCR1156....1D} \\
ZTF20abynbhh & AT\,2021hsv & ATLAS21kna & 112.67908 & 41.46758 & III &  &  &  &  \\
ZTF20achpcvt & AT\,2020vwl & ATLAS20bdgk, Gaia20etp & 232.65758 & 26.98244 & II & TDE & 0.035 &  & \cite{2021TNSCR.159....1H} \\
ZTF20ackzfvn & SN\,2021adcw & ATLAS21blvp, PS21ojv & 31.137 & -21.11931 & VII & SN II & 0.017 &  & \cite{2021TNSAN.297....1R,2021TNSCR4027....1R} \\
ZTF20acnwgtl & SN\,2020yhh &  & 329.42018 & -5.67019 & III & SN Ia & 0.14 &  & \cite{2020TNSAN.216....1G,2020TNSCR3360....1G} \\
ZTF20acobevy & SN\,2020yub & ATLAS20bezd & 0.56296 & -23.49506 & III & SN Ia & 0.08 &  & \cite{2020TNSCR3361....1B} \\
ZTF20acoxmsr & SN\,2020yza & ATLAS20bfpa, PS20mwz & 160.39647 & -3.44147 & III & SN Ia & 0.071 &  & \cite{2020TNSCR3412....1B} \\
ZTF20acpwjus & SN\,2020zjv & ATLAS20bfsh, PS20nfq & 141.63857 & -1.65029 & VI & SN Ia-91T-like & 0.075 &  & \cite{2020TNSCR3486....1I,2020TNSAN.233....1G} \\
ZTF20acpylns & SN\,2020zlc &  & 145.34409 & 15.9613 & III & SN Ia & 0.076388 &  & \cite{2020TNSCR3499....1B} \\
ZTF20acqjmzk & SN\,2020zpj &  & 340.91099 & -14.06795 & VI & SN Ia & 0.1 &  & \cite{2020TNSAN.234....1G,2020TNSCR3524....1I} \\
ZTF20acqzpta & SN\,2020aaez & ATLAS20bgmh & 24.47855 & 30.10691 & III & SN Ia & 0.07135 &  & \cite{2020TNSCR3605....1H,2021TNSCR.512....1D} \\
ZTF20acrheie & SN\,2020aarw &  & 81.31309 & -8.62605 & III & SN Ia & 0.049431 &  & \cite{2020TNSCR3605....1H} \\
ZTF20acsdykp & SN\,2020aaxd & Gaia20fox, PS20lul & 124.86338 & 2.10144 & III & SN Ia & 0.1 &  & \cite{2020TNSCR3605....1H} \\
ZTF20acselme & SN\,2021huw &  & 134.13621 & 39.14126 & III & SN Ia & 0.042 &  & \cite{2021TNSCR1099....1L} \\
ZTF20actdeqw & SN\,2020abfd & ATLAS20bgfy, Gaia20fqk & 26.94368 & -21.76087 & III & SN Ia & 0.045 &  & \cite{2020TNSCR3823....1B} \\
ZTF20actlvpy & SN\,2020aavu & ATLAS20bgwg, PS20lyk & 328.80104 & -11.32326 & III & SN Ia & 0.11 &  & \cite{2020TNSCR3662....1B} \\
ZTF20actrovz & SN\,2020aazy & ATLAS20bgju, PS20nkl & 357.11884 & -5.39563 & III & SN Ia & 0.09 &  & \cite{2020TNSCR3662....1B} \\
ZTF20actvgsy & SN\,2020aazk & ATLAS20bhjy, PS20mhg & 139.69959 & 30.35468 & III & SN Ia & 0.10196 &  & \cite{2020TNSCR3626....1B} \\
ZTF20actvlbq & SN\,2020abaf & ATLAS20bgnd, PS20mnx & 152.56933 & 7.95562 & III & SN Ia & 0.083038 &  & \cite{2020TNSCR3778....1H} \\
ZTF20acuaqlf & SN\,2020abbo & Gaia20fsb & 357.77519 & 6.94249 & III & SN II & 0.017 &  & \cite{2020TNSCR3662....1B,2020TNSCR3790....1D} \\
ZTF20acujaft & SN\,2020abgn & ATLAS20bgmz, Gaia20fsf, PS20mwx & 32.4228 & -6.73706 & III & SN Ia & 0.08 &  & \cite{2020TNSCR3689....1H,2021TNSCR.479....1D} \\
ZTF20aculruc & SN\,2020abij & ATLAS20bgvv & 358.55995 & -4.80556 & III & SN Ia & 0.077 &  & \cite{2020TNSCR3837....1D} \\
ZTF20acumcrz & SN\,2020abln &  & 22.91202 & 18.24501 & VI & SN Ia & 0.054 &  & \cite{2020TNSAN.244....1A,2020TNSCR3712....1P} \\
ZTF20acuoclk & SN\,2020abqd & ATLAS20bgyf, PS20mnz & 124.06977 & 37.40383 & III & SN Ia & 0.082 &  & \cite{2020TNSCR3713....1H} \\
ZTF20acveadu & SN\,2020absk & ATLAS18bcmv, PS20mbv & 136.00384 & -0.08888 & III & SN II & 0.01889 &  & \cite{2020TNSCR3838....1P} \\
ZTF20acvebcu & SN\,2020abqx & ATLAS20bgye, PS20mkm & 178.10281 & 67.5477 & III & SN Ib & 0.063 &  & \cite{2021TNSCR..50....1B} \\
ZTF20acvexen & AT\,2020abrr & ATLAS20bjsp, PS20nwf & 195.44765 & 32.60167 & V &  &  & \tablenotemark{c} &  \\
ZTF20acvjagm & SN\,2020abtf & ATLAS20bgve, PS21akm & 119.89367 & 15.29998 & VI & SN II & 0.014 &  & \cite{2020TNSAN.244....1A,2020TNSCR3712....1P} \\
ZTF20acvkqxy & SN\,2020abve & ATLAS20bjhg & 142.22477 & 40.88916 & III & SN Ia & 0.063327 &  & \cite{2020TNSCR3859....1D} \\
ZTF20acwnrty & SN\,2020acef & ATLAS20bhob, PS20mhi & 146.09988 & 22.97916 & VI & SN Ia-91T-like & 0.065 &  & \cite{2020TNSAN.258....1P,2020TNSCR3811....1P} \\
ZTF20acwofhd & SN\,2020acfp & ATLAS20bjsc & 171.89415 & 47.37943 & III & SN Ic & 0.032816 &  & \cite{2021TNSCR.223....1D,2021TNSCR.358....1D} \\
ZTF20acwpaov & SN\,2020acfq & ATLAS20bjnr & 132.39613 & 9.58738 & III & SN Ia & 0.05913 &  & \cite{2020TNSCR3914....1D} \\
ZTF20acwpixx & SN\,2020aceu & ATLAS20bjqf, PS20nuq & 178.07599 & 7.89155 & II & SN Ia & 0.077451 &  & \cite{2021TNSCR..38....1D} \\
ZTF20acwqmul & AT\,2020acbn & ATLAS20bhzn, ZTF18aatfpbe & 190.1667 & 13.81572 & III &  &  & \tablenotemark{c} &  \\
ZTF20acxdawc & SN\,2020acmf & ATLAS20bigv & 40.86169 & 16.66293 & VI & SN Ia & 0.025 &  & \cite{2020TNSCR3887....1P,2020TNSAN.262....1P} \\
ZTF20acxycfh & SN\,2020acwp &  & 147.03499 & 6.81964 & III & SN Ia & 0.078525 &  & \cite{2020TNSCR3963....1B} \\
ZTF20acynlkj & SN\,2020aden & ATLAS20bkdg & 17.00425 & -15.40552 & III & SN Ia & 0.055008 &  & \cite{2020TNSCR3945....1H} \\
ZTF20acywefl & SN\,2020adka & ATLAS21asd & 118.36714 & 28.26255 & III & SN Ia & 0.058596 &  & \cite{2021TNSCR..66....1D,2021TNSCR..36....1B} \\
ZTF20acyxomd & AT\,2020aeuf & ATLAS20bkau & 150.36432 & 19.46194 & VI &  &  &  &  \\
ZTF20adaftef & SN\,2020adra & ATLAS21atx & 151.3117 & 21.91095 & II & SN Ia & 0.084025 &  & \cite{2021TNSCR..87....1D} \\
ZTF21aaaaddl & SN\,2021D & ATLAS21asf & 358.64203 & 15.74521 & VI & SN Ia & 0.07459 &  & \cite{2021TNSCR.193....1D} \\
ZTF21aaaosmn & SN\,2021ct & ATLAS21axd & 221.76091 & 56.31489 & III & SN Ia & 0.0375 &  & \cite{2021TNSCR.193....1D,2021TNSCR.158....1P} \\
ZTF21aaaovuq & AT\,2021bc & ATLAS21czg & 233.88764 & 18.14889 & II &  &  &  &  \\
ZTF21aaapesv & AT\,2021gk &  & 223.30836 & 28.85542 & III &  &  &  &  \\
ZTF21aaapfal & AT\,2021da & ATLAS21crw & 250.53913 & 35.60025 & II &  &  & \tablenotemark{c} &  \\
ZTF21aaarmti & SN\,2021ek & ATLAS21ajr, PS21fo & 50.95792 & -10.0448 & V & SLSN-I & 0.193 &  & \cite{2021TNSAN..11....1S,2021TNSCR..86....1G} \\
ZTF21aaaytny & SN\,2021mc & ATLAS21cvk & 240.43947 & 18.01928 & II & SN Ia & 0.0459 &  & \cite{2021TNSCR.204....1P} \\
ZTF21aacigst & AT\,2021sq & ATLAS21bsd & 149.86789 & 28.37349 & V &  &  & \tablenotemark{a} &  \\
ZTF21aacnjlf & AT\,2021acw & ATLAS21bnv & 25.84651 & 2.22959 & V &  &  &  &  \\
ZTF21aacudxe & SN\,2021vt & ATLAS21chd, PS21bys & 201.78258 & 11.35239 & V & SN Ia & 0.087 &  & \cite{2021TNSAN..36....1M,2021TNSCR.287....1M} \\
ZTF21aadahyi & SN\,2021akb & ATLAS21cac & 170.90465 & -5.53246 & VI & SN Ia & 0.097 &  & \cite{2021TNSAN..23....1D,2021TNSCR.215....1D} \\
ZTF21aadatfg & SN\,2021xv & ATLAS21cmb & 241.88673 & 36.77951 & II & SN Ic-BL & 0.05 &  & \cite{2021TNSAN..27....1S,2021TNSCR.244....1S} \\
ZTF21aadkgpm & SN\,2021zu & ATLAS21cgz, Gaia21ahe, PS21aqu & 168.74875 & 27.45333 & II & SN Ia & 0.06 &  & \cite{2021TNSCR.223....1D,2021TNSAN..23....1D,2021TNSCR.215....1D} \\
ZTF21aadrmok & AT\,2021acb &  & 261.05295 & 46.82644 & II &  &  & \tablenotemark{c} &  \\
ZTF21aadrmsv & AT\,2021amq & ATLAS21czo & 242.97524 & 21.0087 & V &  &  &  &  \\
ZTF21aadrtcs & SN\,2021abt & ATLAS21cae & 199.82393 & -7.2492 & II & SN Ia & 0.0014285 &  & \cite{2021TNSCR.268....1G} \\
ZTF21aadrtqz & SN\,2021abn & ATLAS21bmj & 208.7036 & -13.60863 & II & SN Ia & 0.058 &  & \cite{2021TNSCR.250....1H} \\
ZTF21aaekkbv & SN\,2021ait & ATLAS21crv, Gaia21anh & 264.64777 & 42.21554 & II & SN Ia & 0.07 &  & \cite{2021TNSCR.277....1S} \\
ZTF21aaekmoy & SN\,2021ais & ATLAS21cry & 223.53354 & 2.97951 & II & SN Ia & 0.080003 &  & \cite{2021TNSCR.306....1H} \\
ZTF21aaevrjl & SN\,2021arg & ATLAS21dkx, PS21aia & 67.82827 & -10.39637 & V & SN II & 0.031 &  & \cite{2021TNSAN..29....1G,2021TNSCR.230....1G,2021TNSCR.259....1G,2021TNSAN..25....1G} \\
ZTF21aafdvxz & SN\,2021bkb & ATLAS21dhf, PS21ewl & 236.87341 & 10.93365 & III & SN II & 0.05 &  & \cite{2021TNSCR.405....1D} \\
ZTF21aafkktu & SN\,2021avg & ATLAS21cux, Gaia21aku, PS21bkp & 174.99587 & 14.52796 & II & SN II & 0.031 &  & \cite{2021TNSAN..27....1S,2021TNSCR.244....1S} \\
ZTF21aaflcbk & SN\,2021arr & ATLAS21dab, Gaia21aml, PS21bzh & 189.12726 & 19.82585 & VI & SN Ia & 0.07 &  & \cite{2021TNSCR.405....1D} \\
ZTF21aaglrzc & SN\,2021bmc & ATLAS21djm & 127.24882 & 12.41938 & II & SN II & 0.05 &  & \cite{2021TNSCR.327....1M,2021TNSAN..50....1T,2021TNSCR1378....1D} \\
ZTF21aagnbsu & AT\,2021cld & ATLAS21efr & 124.66522 & 6.12749 & V &  &  &  &  \\
ZTF21aagrzqz & AT\,2021chg (also AT\,2013kz) & ATLAS21ehz & 144.49052 & 48.3884 & II &  &  & \tablenotemark{c} &  \\
ZTF21aagsbot & AT\,2021bvu & ATLAS21emi & 158.1223 & 16.16226 & V &  &  &  &  \\
ZTF21aagshha & AT\,2021byg & ATLAS21emb & 171.75836 & 16.73231 & III &  &  & \tablenotemark{c} &  \\
ZTF21aagskhr & AT\,2021bpz & ATLAS21end & 191.98612 & 29.72008 & V &  &  & \tablenotemark{a} &  \\
ZTF21aagtcgb & SN\,2021bmb & ATLAS21djl, PS21zq & 224.32537 & 43.45972 & II & SN Ia & 0.0375 &  & \cite{2021TNSCR.405....1D} \\
ZTF21aagtexi & SN\,2021bqo & ATLAS21eie & 188.74309 & 45.44921 & II & SN Ia & 0.08 &  & \cite{2021TNSCR.435....1D} \\
ZTF21aagtquu & AT\,2021brg & ATLAS21eem & 244.44327 & 65.12367 & III &  &  & \tablenotemark{c} &  \\
ZTF21aagxmcs & SN\,2021cca & ATLAS21efx, PS21zy & 154.0805 & -1.78745 & V & SN Ia & 0.085 &  & \cite{2021TNSCR.434....1D} \\
ZTF21aagzswk &  &  & 26.91632 & -1.97598 & VI &  &  &  &  \\
ZTF21aagzwod &  &  & 19.41428 & -2.94091 & III &  &  &  &  \\
ZTF21aahfizx & AT\,2021bwg & ATLAS21egb & 225.31005 & 13.0221 & V &  &  &  &  \\
ZTF21aahfjbs & SN\,2021cky & ATLAS21eev & 229.39754 & 18.11804 & V & SN Ia-pec & 0.073 &  & \cite{2021TNSCR.446....1P,2021TNSAN..61....1P} \\
ZTF21aahfjlo & SN\,2021clz & ATLAS21flj, Gaia21bfk, PS21axb & 214.64922 & 52.1011 & II & SN Ia & 0.039 &  & \cite{2021TNSCR.479....1D} \\
ZTF21aahhmev & AT\,2021bzb & ATLAS21els & 76.697 & 3.75387 & V &  &  &  &  \\
ZTF21aahnkut & SN\,2021cez & ATLAS21epx & 197.81538 & -16.54579 & II & SN Ia & 0.07 &  & \cite{2021TNSCR.569....1D} \\
ZTF21aahnqnn & AT\,2021dng & ATLAS21gvh & 167.32333 & -5.75498 & V &  &  &  &  \\
ZTF21aahpoxv & AT\,2021csz & ATLAS21gau, PS21cqr & 174.10466 & 4.8955 & V &  &  &  &  \\
ZTF21aahpxww & AT\,2021cff & ATLAS21fwh & 158.95735 & 10.91143 & VI &  &  &  &  \\
ZTF21aahzspd & SN\,2021cjx & ATLAS21fkc & 148.00477 & 21.17413 & V & SN Ia & 0.11 &  & \cite{2021TNSCR.569....1D} \\
ZTF21aaiucta & AT\,2021cfu & ATLAS21fza, PS21cnz & 254.84455 & -1.20244 & VI &  &  &  &  \\
ZTF21aaiucta & AT\,2021cfu & ATLAS21fza, PS21cnz & 254.84454 & -1.20245 & III &  &  & \tablenotemark{c} &  \\
ZTF21aajgdeu & SN\,2021cjd & ATLAS21ekw & 193.27745 & 36.81955 & III & SN IIP & 0.027929 &  & \cite{2021TNSCR1234....1D} \\
ZTF21aajgdpo & SN\,2021cjy & ATLAS21hdg, PS21aei & 194.82941 & 38.88605 & II & SN II & 0.036088001 &  & \cite{2021TNSCR1664....1D} \\
ZTF21aakbgpf & SN\,2021crv & ATLAS21gdz, PS21bhh & 122.63426 & -8.12558 & II & SN Ia & 0.05 &  & \cite{2021TNSCR.650....1D} \\
ZTF21aakfqwq & AT\,2021crk & ATLAS21hdc, PS21cpw & 176.27889 & 18.54037 & V &  &  &  &  \\
ZTF21aalgboj & AT\,2021dfo & ATLAS21hcp & 135.79646 & 51.78792 & II &  &  & \tablenotemark{c} &  \\
ZTF21aalgqsv & SN\,2021dic & PS21bxl & 154.08238 & 0.65215 & II & SN Ia & 0.079358 &  & \cite{2021TNSCR.665....1D} \\
ZTF21aalhgqi & SN\,2021dib & ATLAS21gtb, PS21aie & 173.7361 & 9.03865 & VI & SN I & 0.086 &  & \cite{2021TNSAN..81....1M,2021TNSCR.679....1M} \\
ZTF21aalnxkl & SN\,2021djz & PS21bze & 202.98122 & -2.45151 & V & SN Ia & 0.087 &  & \cite{2021TNSCR.563....1H,2021TNSAN..72....1H} \\
ZTF21aalxxzn & AT\,2021fxu & Gaia22dgm & 206.82834 & 2.18269 & VII & AGN & 0.1097 &  & \cite{2022TNSCR1320....1N} \\
ZTF21aamjzki & AT\,2020hoj & ZTF20aavfogu & 173.45805 & -20.81437 & III &  &  &  &  \\
ZTF21aamkxbl & SN\,2021dwo & ATLAS21gwk & 216.18103 & -1.57031 & II & SN Ia & 0.057153 &  & \cite{2021TNSCR.680....1B} \\
ZTF21aamucom & SN\,2021dsb & ATLAS21hgi, Gaia21byu, PS21bxv & 158.82831 & -13.00542 & II & SN Ia-91T-like & 0.025 &  & \cite{2021TNSCR.665....1D} \\
ZTF21aamxduf & SN\,2021eij & ATLAS21hjg & 217.38808 & 10.94154 & VI & SN Ia & 0.1 &  & \cite{2021TNSAN..81....1M,2021TNSCR.679....1M} \\
ZTF21aanswls & SN\,2021ell & ATLAS19bftq & 298.58614 & 1.58498 & III & SN Ia & 0.025 &  & \cite{2021TNSCR.953....1H} \\
ZTF21aanuyro & SN\,2021eno & ATLAS21huk, Gaia21blr, PS21ctb & 224.82649 & 59.90032 & II & SN Ia & 0.07 &  & \cite{2021TNSCR1008....1D} \\
ZTF21aanvtng & AT\,2021ehx &  & 167.89582 & 5.58533 & V &  &  &  &  \\
ZTF21aanxhjv & AT\,2021ehb & ATLAS21jdy & 46.94925 & 40.3113 & II & TDE & 0.018 &  & \cite{2021TNSAN.103....1G,2021TNSCR2295....1Y,2021TNSAN.183....1Y,2021TNSAN.309....1A,2022TNSCR2102....1Y} \\
ZTF21aaocgci & AT\,2021eox & ATLAS21hnu & 128.46705 & -11.86269 & II &  &  & \tablenotemark{c} &  \\
ZTF21aaocibg & SN\,2021epo & ATLAS21hpb, Gaia21biq, PS21cgw & 131.54246 & 56.12765 & II & SN Ia & 0.071 &  & \cite{2021TNSCR.885....1D} \\
ZTF21aaoexjt & SN\,2021ezs & ATLAS21hsb & 208.39581 & 27.08093 & II & SN Ia & 0.11 &  & \cite{2021TNSCR.922....1D} \\
ZTF21aaomiwf &  &  & 194.63988 & 83.11293 & III &  &  &  &  \\
ZTF21aaootdm & AT\,2021gny (also AT\,2015dv) & ATLAS21ith, PS21cgk & 211.95257 & 7.1261 & V &  &  &  &  \\
ZTF21aaopupn & AT\,2021fap & ATLAS21hon & 217.35744 & 9.06336 & V &  &  &  &  \\
ZTF21aaoqayg & AT\,2021eyw & ATLAS21iuk & 252.98655 & 28.92162 & III &  &  & \tablenotemark{c} &  \\
ZTF21aaoqcnh & SN\,2021feu & ATLAS21ixy & 272.47882 & 11.30154 & III & SN Ia & 0.04 &  & \cite{2021TNSCR.762....1K,2021TNSAN..84....1K} \\
ZTF21aaovdlh & SN\,2021fck & ATLAS21huq, PS21bxe & 130.34449 & 2.88978 & II & SN Ia & 0.056 &  & \cite{2021TNSCR.885....1D} \\
ZTF21aapfmut & AT\,2021fyr & ATLAS21ikk, PS21fzt & 154.93818 & 32.24469 & III &  &  & \tablenotemark{c} &  \\
ZTF21aaphjwb & AT\,2021ghw & ATLAS21kdb & 177.84396 & 24.65899 & V &  &  &  &  \\
ZTF21aaphzor & AT\,2021hcd &  & 154.5871 & 5.92311 & VI &  &  &  &  \\
ZTF21aapjmgf & SN\,2021fzp & ATLAS21iif, Gaia21bkh, PS21clp & 220.98371 & 16.41324 & II & SN II & 0.054 &  & \cite{2021TNSCR.900....1W,2021TNSAN..98....1W,2021TNSCR1234....1D} \\
ZTF21aapjmpq & SN\,2021fzq & ATLAS21imm, PS21cgr & 225.28036 & 12.31399 & VI & SN Ia-91T-like & 0.13 &  & \cite{2021TNSAN.111....1P,2021TNSCR1066....1P} \\
ZTF21aapjmpq & SN\,2021fzq & ATLAS21imm, PS21cgr & 225.28036 & 12.31399 & VI & SN Ia-91T-like & 0.13 &  & \cite{2021TNSAN.111....1P,2021TNSCR1066....1P} \\
ZTF21aapjpee & AT\,2021hdo & ATLAS21jdj & 202.63239 & -6.3692 & V &  &  &  &  \\
ZTF21aapkuur & SN\,2021gba & ATLAS21iuc, PS21bxj & 219.26622 & -8.07644 & V & SN Ia & 0.1 &  & \cite{2021TNSCR1251....1D,2021TNSAN.126....1D} \\
ZTF21aappehx & SN\,2021gwu &  & 139.10441 & -17.35608 & II & SN Ia & 0.054514 &  & \cite{2021TNSCR.953....1H} \\
ZTF21aaprfkc & SN\,2021gwp & ATLAS21jcn, PS21ego & 199.01655 & -15.50397 & III & SN Ia & 0.056 &  & \cite{2021TNSCR1197....1D} \\
ZTF21aaptmrn & SN\,2021gdm & ATLAS21itu & 237.38765 & 4.67097 & VI & SN Ia & 0.112 &  & \cite{2021TNSAN.118....1P,2021TNSCR1178....1P} \\
ZTF21aapuwry & AT\,2021gdw & PS21clx & 249.19601 & -16.22766 & VI &  &  &  &  \\
ZTF21aapvxnf & SN\,2021gyw & ATLAS21kpq & 281.88222 & 64.46884 & II & SN Ia-91T-like & 0.085 &  & \cite{2021TNSCR.965....1D} \\
ZTF21aapxogj & SN\,2021ghj & PS21cfn & 125.95185 & 38.38982 & II & SN II & 0.060214 &  & \cite{2021TNSCR1574....1D} \\
ZTF21aaqgrrf & SN\,2021gqm & ATLAS21jlu, PS21eeh & 153.76115 & 3.79654 & V & SN Ia & 0.082 &  & \cite{2021TNSAN.110....1M,2021TNSCR1045....1P} \\
ZTF21aaqhqke & SN\,2021gpw & ATLAS21iwv, Gaia21czj & 200.47858 & 16.74437 & II & SN IIn & 0.075 &  & \cite{2021TNSCR1071....1P} \\
ZTF21aaqjuyt &  &  & 165.70851 & 23.85029 & VI &  &  &  &  \\
ZTF21aaqmoof & SN\,2021gvv & ATLAS21jlj, PS21fbg & 256.14998 & 14.08037 & V & SN II & 0.038 &  & \cite{2021TNSCR1155....1P,2021TNSAN.116....1P} \\
ZTF21aaqpykm & SN\,2021hay & ATLAS21jli, PS21gdb & 229.5394 & 15.17631 & V & SN Ia & 0.06 &  & \cite{2021TNSCR1046....1P} \\
ZTF21aaqwbgq & SN\,2021hjb & ATLAS21jgx & 133.93957 & 16.92466 & III & SN Ia & 0.06651 &  & \cite{2021TNSCR1102....1D} \\
ZTF21aarigsr & SN\,2021iaf & ATLAS21kat, Gaia21chu & 227.84385 & 26.79227 & II & SN Ia & 0.059 &  & \cite{2021TNSCR1283....1S} \\
ZTF21aarmuxl & AT\,2021hpp & ATLAS21kne, PS21cus & 141.24411 & 42.80907 & III &  &  & \tablenotemark{c} &  \\
ZTF21aarohyu & SN\,2021hzo & ATLAS21khv & 180.16705 & 51.84093 & II & SN Ia & 0.063 &  & \cite{2021TNSCR1147....1D} \\
ZTF21aarrkwn & AT\,2021igm & ATLAS21kdi, PS21fxj & 226.85561 & 18.31667 & V &  &  &  &  \\
ZTF21aarteuc & AT\,2021hzc & ATLAS21kxo, PS21dfc & 151.98128 & -8.28587 & V &  &  &  &  \\
ZTF21aarycyl & SN\,2021hvu & ATLAS21jpb, Gaia21bxb, PS21fia & 153.41 & -24.52076 & II & SN Ia-91T-like & 0.104 &  & \cite{2021TNSCR1305....1H} \\
ZTF21aasgcve & SN\,2021idn & ATLAS21kye & 129.16748 & 1.16559 & V & SN Ia-91T-like & 0.09 &  & \cite{2021TNSCR1282....1D} \\
ZTF21aasgrpw & SN\,2021idh & ATLAS21laz, PS21dhx & 155.82763 & -0.42939 & VI & SN Ia & 0.095 &  & \cite{2021TNSCR1224....1D} \\
ZTF21aaskuth & AT\,2021icw & ATLAS21lge, PS21fxn & 262.75059 & 7.96276 & VI &  &  &  &  \\
ZTF21aassamj & SN\,2021ify & ATLAS21lot, PS21fig & 261.72373 & 13.17885 & II & SN Ia & 0.056 &  & \cite{2021TNSCR1155....1P,2021TNSCR1197....1D,2021TNSAN.116....1P} \\
ZTF21aaswvyc & SN\,2021ijy & ATLAS21lyk & 258.4873 & 39.0988 & III & SN Ia & 0.07 &  & \cite{2021TNSCR1198....1T} \\
ZTF21aatdzmt & SN\,2021imq & ATLAS21lqa, Gaia21cfl, PS21fih & 227.36531 & 36.81745 & II & SN Ia & 0.076 &  & \cite{2021TNSCR1252....1S} \\
ZTF21aatlbsi & SN\,2021ipy & ATLAS21ncf & 292.34717 & 74.07815 & II & SN Ia & 0.07 &  & \cite{2021TNSCR1197....1D} \\
ZTF21aatyxox & AT\,2021jtw & ATLAS21mtq, PS21fwq & 206.9273 & -26.37541 & V &  &  &  &  \\
ZTF21aauufbz &  &  & 160.2291 & 24.74699 & IV &  &  & \tablenotemark{a} &  \\
ZTF21aauufgo & AT\,2021jjp & ATLAS21mwb, PS21eik & 211.12748 & -24.86573 & V &  &  &  &  \\
ZTF21aauurqh &  &  & 204.13567 & -5.05061 & VI &  &  &  &  \\
ZTF21aavpgvx & AT\,2021jui & ATLAS21mwc, PS21fnl & 186.71367 & -17.77826 & V &  &  &  &  \\
ZTF21aavrywc & AT\,2021jwa & ATLAS21njq & 236.35534 & 8.20475 & V &  &  &  &  \\
ZTF21aawaguv & AT\,2021kae &  & 181.57238 & 5.81541 & V &  &  &  &  \\
ZTF21aawckpe &  &  & 198.85353 & 1.12187 & III &  &  &  &  \\
ZTF21aawtazf & SN\,2021kkz & ATLAS21nnp, PS21gar & 248.98476 & 18.08285 & III & SN Ia & 0.061948 &  & \cite{2021TNSCR1482....1D,2021TNSCR1460....1J} \\
ZTF21aaxjiii & SN\,2021klq & ATLAS21nqa & 351.50055 & 46.65778 & III & SN Ia & 0.04 &  & \cite{2021TNSCR2075....1B} \\
ZTF21aaxkckg & SN\,2021kmv & ATLAS21nme, PS21elb & 152.84195 & 41.78835 & II & SN Ia & 0.063 &  & \cite{2021TNSCR1464....1S} \\
ZTF21aaxstsv & SN\,2021ktw & ATLAS21nqy & 215.87831 & 8.34491 & II & SN Ia & 0.09 &  & \cite{2021TNSCR1592....1S} \\
ZTF21aaxtpty & AT\,2021kqp & ATLAS21oaf, PS21doo & 223.79475 & -6.98593 & V &  &  &  &  \\
ZTF21aaxxihx & SN\,2021ktv & ATLAS21nqw & 165.76618 & 8.86105 & III & SN Ic-BL & 0.06 &  & \cite{2021TNSCR1718....1T} \\
ZTF21aaxxjdr & AT\,2021ktu & ATLAS21nqv & 169.0091 & 21.29421 & II &  &  &  &  \\
ZTF21aaxxjen & SN\,2021kun & ATLAS21ofj & 166.41407 & 19.46031 & I & SN Ia & 0.097 &  & \cite{2021TNSCR1631....1G} \\
ZTF21aaxyecd & SN\,2021kui & ATLAS21ohm & 194.25355 & 40.43447 & II & SN Ia & 0.061249 &  & \cite{2021TNSCR1575....1S} \\
ZTF21aaydomp & AT\,2021ldb & ATLAS21oan & 190.5094 & -14.12457 & VI &  &  &  &  \\
ZTF21aaydtqk & AT\,2021kxh & ATLAS21owr & 178.90335 & 5.54756 & V &  &  &  &  \\
ZTF21aayoiis & AT\,2021mpc & ATLAS21pes, PS21gct & 179.09923 & -17.53295 & V &  &  &  &  \\
ZTF21aayqrgx & SN\,2021lax & Gaia21cof & 212.33849 & 72.11727 & II & SN Ib & 0.033 &  & \cite{} \\
ZTF21aazlxiw & SN\,2021lta & ATLAS21oez, PS21dqe & 246.50245 & 28.42744 & II & SN Ia & 0.1 &  & \cite{2021TNSCR1836....1S} \\
ZTF21aazmjaf & AT\,2021lkq & ATLAS21ovp & 316.71025 & 10.73573 & V &  &  &  &  \\
ZTF21aazpyza & AT\,2021lob &  & 222.37709 & -27.24104 & V &  &  &  &  \\
ZTF21aazpzqo & AT\,2021loa &  & 214.60334 & -27.65508 & III &  &  &  &  \\
ZTF21aazqtor & SN\,2021lny & ATLAS21oyt & 205.26646 & -25.60261 & II & SN Ia & 0.0954 &  & \cite{2021TNSCR1717....1H} \\
ZTF21aazqynq & AT\,2021lsw & ATLAS21osi & 186.93889 & 26.77057 & V &  &  &  &  \\
ZTF21aazybsh &  &  & 16.08828 & 55.74215 & III &  &  &  &  \\
ZTF21abaagct & AT\,2021lrw & ATLAS21orf, PS21gdn & 305.75781 & -6.41894 & V &  &  &  &  \\
ZTF21abalkop & AT\,2021mho & ATLAS21rcr & 183.01062 & 19.04756 & VI &  &  &  &  \\
ZTF21abamnyi & AT\,2021mck & ATLAS21rch & 220.04695 & 66.13794 & V &  &  &  &  \\
ZTF21abawvgt & AT\,2021mbe & ATLAS21pjw & 326.67805 & 8.54946 & VI &  &  &  &  \\
ZTF21abbliav & SN\,2021mqv & ATLAS21peg, Gaia21cte & 322.88652 & 11.14894 & VI & SN Ia & 0.049 &  & \cite{2021TNSCR1740....1S} \\
ZTF21abboaaz & AT\,2021mkx & ATLAS21qbe & 179.52419 & -27.75361 & III &  &  &  &  \\
ZTF21abbuxzr & AT\,2021mmq & ATLAS21pad & 167.3917 & -1.5678 & III & Varstar &  &  & \cite{2021TNSCR1859....1H} \\
ZTF21abbxdcm & SN\,2021msu & ATLAS21pis & 223.63703 & 9.22811 & II & SN Ia-91T-like & 0.05 &  & \cite{2021TNSCR1812....1N,2021TNSCR1877....1N} \\
ZTF21abbxlbu & AT\,2021mqq & ATLAS21pju & 315.86938 & 4.94065 & II &  &  &  &  \\
ZTF21abbytmm & SN\,2021msv & ATLAS21piv & 195.09912 & -11.9655 & II & SN Ia & 0.054 &  & \cite{2021TNSCR1861....1S} \\
ZTF21abbzjeq & SN\,2021mwb & ATLAS21pjy, PS21dxq & 242.23797 & 35.42108 & II & SN Ia & 0.043 &  & \cite{} \\
ZTF21abcfgwv & AT\,2021mxu & ATLAS21rfm & 171.34341 & 24.90104 & V &  &  &  &  \\
ZTF21abcixor & SN\,2021nip & ATLAS21prb, PS21ety & 244.21006 & 12.64293 & II & SN II & 0.033 &  & \cite{2021TNSCR1984....1D,2021TNSCR1952....1P} \\
ZTF21abcmepi & SN\,2021nlj & ATLAS21pvm, PS21hrs & 187.86256 & 4.22648 & II & SN Ia & 0.089 &  & \cite{2021TNSCR1939....1S} \\
ZTF21abcrhpk & SN\,2021nqo & ATLAS21pvj, PS21gzv & 322.36694 & 4.8875 & II & SN Ia & 0.0744 &  & \cite{2021TNSCR2128....1T,2021TNSCR2140....1S} \\
ZTF21abdbodg & SN\,2021ohp & Gaia21czd & 206.99498 & -13.78255 & III & SN Ia & 0.064 &  & \cite{2021TNSCR2021....1S} \\
ZTF21abdmevk & SN\,2021ont & ATLAS21qky, PS21ivp & 246.67702 & 39.14524 & II & SN II & 0.02833 &  & \cite{2021TNSAN.170....1M,2021TNSCR1964....1M,2021TNSAN.200....1B,2021TNSCR2547....1C} \\
ZTF21abfoyac & SN\,2021pni & Gaia21ebi & 268.07435 & 8.23976 & III & SN II & 0.033 &  & \cite{2021TNSCR2152....1T,2021TNSCR2547....1C} \\
ZTF21abfxibf & SN\,2021qtn &  & 355.38227 & 14.12417 & II & SN Ia & 0.063044 &  & \cite{2021TNSCR4017....1G} \\
ZTF21abgxbvi & AT\,2021quh &  & 250.76254 & -26.69082 & III &  &  &  &  \\
ZTF21abhhbmd & SN\,2021qnv & ATLAS21wwc, Gaia21dox & 338.16656 & 28.20678 & III & SN Ia & 0.042 &  & \cite{2021TNSCR2287....1H,2021TNSCR2547....1C} \\
ZTF21abhshmt & SN\,2021qvg & ATLAS21ztt, Gaia21dpf & 200.08813 & 48.72436 & II & SN Ia-91T-like & 0.055 &  & \cite{2021TNSCR2287....1H} \\
ZTF21abhuoyz & SN\,2021qyh & ATLAS21baec, PS21hwf & 215.89982 & 14.11976 & II & SN Ia & 0.0875 &  & \cite{2021TNSCR2303....1H} \\
ZTF21abhymom & SN\,2021qyf & ATLAS21zvh & 205.34227 & 1.98462 & III & SN Ia & 0.08 &  & \cite{2021TNSCR2380....1N} \\
ZTF21abhzboh & SN\,2021qyc & ATLAS21zwq & 237.23744 & 6.84263 & II & SN Ia & 0.051097 &  & \cite{2021TNSCR2303....1H} \\
ZTF21abidtrd & SN\,2021rax & ATLAS21baeg & 283.24974 & 28.95966 & III & SN Ia & 0.048 &  & \cite{2021TNSCR2317....1H} \\
ZTF21abifutc & SN\,2021rdq & ATLAS21baen & 190.19495 & 42.30288 & II & SN Ia & 0.072 &  & \cite{2021TNSCR2369....1S} \\
ZTF21abjbjba & SN\,2021sbv & ATLAS21bbfc & 196.47762 & 2.14131 & II & SN Ia & 0.069122 &  & \cite{2021TNSCR2575....1G} \\
ZTF21abjciua & AT\,2021seu & ATLAS21bbfi & 215.39427 & 37.90969 & II & BFF & 0.06 &  & \cite{2021TNSAN.195....1A,2021TNSCR2460....1A} \\
ZTF21abkinhh & SN\,2016elg & iPTF16elg & 212.641 & -0.98848 & III & SN Ia & 0.054027 &  & \cite{2021TNSCR2576....1G} \\
ZTF21abkqkon & SN\,2021swl & ATLAS21bceo, Gaia21djx & 345.65222 & 14.86495 & II & SN Ia & 0.0885 &  & \cite{2021TNSCR2504....1H} \\
ZTF21abmgxyu & SN\,2021tnd & ATLAS21bbvf, PS21imd & 327.23818 & -9.07113 & III & SN Ia & 0.085 &  & \cite{2021TNSCR2591....1B,2021TNSCR2635....1S} \\
ZTF21abmwzxt & SN\,2021tsz & ATLAS21bccc, Gaia21dmi, PS21ikp & 354.49331 & -0.4415 & II & SN II & 0.04 &  & \cite{2021TNSCR2659....1P,2021TNSAN.209....1P} \\
ZTF21abnfgff & SN\,2021tvc & ATLAS21bdey, Gaia21dpc & 350.19705 & 4.261 & III & SN Ia & 0.06 &  & \cite{2021TNSCR2577....1B} \\
ZTF21abnhnxu & AT\,2021uvr & ATLAS21bdyq, PS21iom & 351.75532 & -11.02135 & II &  &  &  &  \\
ZTF21abnlhxs & SN\,2021tyw & ATLAS21bche, Gaia21doz, PS21isy & 346.48525 & 14.35774 & II & SN II & 0.013 &  & \cite{2021TNSAN.206....1D,2021TNSCR2619....1M} \\
ZTF21abnlnmq & AT\,2021txf & ATLAS21beii & 14.57232 & 20.59191 & V &  &  &  &  \\
ZTF21abnvsic & SN\,2021uby & ATLAS21bcgu & 257.17946 & 39.53526 & II & SN Ia & 0.072 &  & \cite{2021TNSCR2624....1S} \\
ZTF21abotose & SN\,2021ugl & ATLAS21bebk & 246.98203 & 20.25316 & VI & SN IIb & 0.0412 &  & \cite{2021TNSCR2834....1C,2021TNSCR2795....1R,2021TNSAN.220....1R} \\
ZTF21abowqqa & SN\,2021uga & ATLAS21bdru & 247.81439 & 12.32168 & V & SN Ia & 0.09 &  & \cite{2021TNSCR2696....1P,2021TNSAN.213....1P} \\
ZTF21abpjpxr & SN\,2021uwa & ATLAS21bdqp & 335.92088 & 19.73504 & V & SN Ia & 0.09 &  & \cite{2021TNSCR2730....1S} \\
ZTF21abpjrqx & AT\,2021uhr & ATLAS21bebe & 321.71605 & 5.75134 & VI &  &  &  &  \\
ZTF21abqfpsc & SN\,2021umz (also AT\,2016jkb) & ATLAS21bdjn, Gaia21ecj & 291.64365 & -26.34741 & V & SN Ia-91T-like & 0.076 &  & \cite{2021TNSCR2710....1N,2021TNSAN.214....1N} \\
ZTF21abqjjrb & SN\,2021uqu & ATLAS21bdzp, PS21inv & 357.29406 & -11.25516 & III & SN & 0.078 &  & \cite{2021TNSCR2952....1T,2021TNSAN.227....1T} \\
ZTF21abrfvax & SN\,2021vis & ATLAS21bfiy & 226.27859 & 21.27113 & VI & SN Ic & 0.0518 &  & \cite{2021TNSCR2980....1C,2021TNSCR3202....1C} \\
ZTF21abrfvax & SN\,2021vis & ATLAS21bfiy & 226.27859 & 21.27113 & III & SN Ic & 0.0518 &  & \cite{2021TNSCR2980....1C,2021TNSCR3202....1C} \\
ZTF21abrotmc & SN\,2021vku & ATLAS21bfzs, PS21jdw & 322.38702 & 8.07912 & V & SN Ia & 0.097 &  & \cite{2021TNSCR2856....1S} \\
ZTF21abrskvb & SN\,2021vkr & ATLAS21bgbc & 308.9457 & 5.45901 & VI & SN Ia & 0.083 &  & \cite{2021TNSCR2892....1S} \\
ZTF21abrzeif & AT\,2021vjy & ATLAS21bgnx & 29.30134 & -4.87903 & VI &  &  &  &  \\
ZTF21absbwyz & SN\,2021vou & ATLAS21bfjz, PS21izi & 249.47325 & 11.59448 & V & SN Ia & 0.09 &  & \cite{2021TNSCR2952....1T,2021TNSCR2980....1C,2021TNSAN.227....1T} \\
ZTF21absjryg & AT\,2021vxq &  & 15.10958 & -4.1558 & VI &  &  &  &  \\
ZTF21abtmsie & AT\,2021wfm & ATLAS21bgln & 339.79123 & 41.78092 & III &  &  &  &  \\
ZTF21abtsoky & SN\,2021zfo & ATLAS21bjdz, PS21kff & 270.11269 & 69.16242 & V & SN Ia & 0.085 &  & \cite{2021TNSAN.251....1N,2021TNSCR3360....1N,2021TNSCR3376....1S} \\
ZTF21abtuutw & SN\,2021whj & ATLAS21bgnp, PS21jyt & 2.30267 & -5.93881 & III & SN Ia & 0.079 &  & \cite{2021TNSCR2979....1G} \\
ZTF21abvyczt & AT\,2021xco & ATLAS21bibd & 247.27735 & -8.48109 & VI &  &  &  &  \\
ZTF21abwycli & SN\,2021xyh & ATLAS21biiw, PS21jtu & 339.96861 & 12.58475 & V & SN II & 0.035 &  & \cite{2021TNSCR3202....1C} \\
ZTF21abxqkjd & AT\,2021xvw & ATLAS21bhyv, PS21nmn & 343.22562 & -19.7127 & V &  &  &  &  \\
ZTF21abxzzys & SN\,2021ygf & ATLAS21binx & 248.96119 & 38.37612 & V & SN Ia & 0.098876 &  & \cite{2021TNSCR3211....1C} \\
ZTF21abyonuw & SN\,2021yip & ATLAS21bjhu, PS21ket & 345.43594 & 28.46884 & VI & SN Ia & 0.128 &  & \cite{2021TNSCR3326....1N} \\
ZTF21abzbuvf & SN\,2021yik & ATLAS21bipy, PS21kvq & 7.50773 & -3.03131 & VI & SN Ia & 0.1 &  & \cite{2021TNSCR3202....1C} \\
ZTF21acaohho & AT\,2021ynv &  & 44.19692 & 0.73207 & V &  &  &  &  \\
ZTF21acbgcma & SN\,2021yvh & ATLAS21bizm, PS21okm & 249.94553 & 32.57075 & III & SN Ia & 0.05239 &  & \cite{2021TNSCR3250....1B} \\
ZTF21acbjxhr & SN\,2021ysn & ATLAS21bjhq, PS21kui & 359.18144 & 13.34985 & V & SN Ia & 0.085 &  & \cite{2021TNSCR3350....1S} \\
ZTF21acbqjsz & AT\,2021zda &  & 312.98109 & 0.09868 & V &  &  &  &  \\
ZTF21acdnnfg & SN\,2021zgu & ATLAS21bjeq, Gaia21end, PS21mgp & 73.43527 & 0.82282 & V & SN Ia & 0.068 &  & \cite{2021TNSCR3394....1S} \\
ZTF21acdozee & AT\,2021zhq & ATLAS21bjcv & 246.67356 & 24.92438 & V &  &  &  &  \\
ZTF21acehvay & SN\,2021aaiq &  & 327.87377 & 26.48246 & III & SN Ia & 0.055291 &  & \cite{2021TNSCR3508....1P} \\
ZTF21acftror & SN\,2021acej &  & 304.47809 & 3.23255 & III & SN Ia & 0.05 &  & \cite{2021TNSCR3640....1N} \\
ZTF21acgylas & SN\,2021abaq & ATLAS21bkml, PS21lui & 347.01665 & 22.57312 & VI & SN Ia & 0.078 &  & \cite{2021TNSCR3694....1G,2021TNSAN.277....1G} \\
ZTF21acgyzel & SN\,2021abam & ATLAS21bkdm, PS21lji & 17.05099 & -19.79002 & II & SN Ia & 0.049 &  & \cite{2021TNSCR3666....1J,2021TNSAN.272....1J} \\
ZTF21acgzxbn & AT\,2021able &  & 73.49642 & -16.38124 & V &  &  &  &  \\
ZTF21acgzytq & AT\,2021abjk & ATLAS21blzh & 84.14974 & -20.07543 & V &  &  &  &  \\
ZTF21achafji & AT\,2021abwc &  & 72.66684 & -20.16099 & VI & Galaxy & 0.089 &  & \cite{2021TNSCR3661....1I,2021TNSAN.274....1I} \\
ZTF21achaxmt & SN\,2021abnc & ATLAS21bkcv & 284.06308 & 33.0193 & II & SN Ia & 0.08 &  & \cite{2021TNSCR3609....1G} \\
ZTF21achbgdo & SN\,2021abbm &  & 282.58432 & 70.48222 & III & SN Ia & 0.089 &  & \cite{2021TNSCR3663....1N} \\
ZTF21achctlk & AT\,2021abdx & PS21ltv & 23.46599 & -19.67035 & V &  &  &  &  \\
ZTF21achdsoo & SN\,2021abmr &  & 6.32401 & -4.15696 & VI & SN Ia & 0.109 &  & \cite{2021TNSCR3674....1I,2021TNSAN.275....1I} \\
ZTF21achdvyu & SN\,2021aadc & ATLAS21blak, PS21kpm & 1.13614 & 19.76144 & VI & SLSN-II & 0.1953 &  & \cite{2021TNSAN.280....1P,2021TNSCR3822....1G} \\
ZTF21achlxxo & SN\,2021abll &  & 71.84171 & 74.02835 & III & SN Ia & 0.08145 &  & \cite{2021TNSCR3663....1N} \\
ZTF21achqiue & SN\,2021abmy & ATLAS21bkiu, Gaia21fca, PS21lta & 355.12566 & 12.93151 & II & SN Ia & 0.028 &  & \cite{2021TNSCR3640....1N} \\
ZTF21achutuk & SN\,2021abpw & ATLAS21bkjt, Gaia21fhj & 311.27769 & -25.73199 & II & SN Ia & 0.063 &  & \cite{2021TNSCR3724....1N} \\
ZTF21achxzrj & SN\,2021abpo & ATLAS21bkzi & 63.27148 & -5.44853 & VI & SN Ia & 0.1 &  & \cite{2021TNSCR3661....1I,2021TNSAN.274....1I} \\
ZTF21achzfpg & SN\,2021absx & ATLAS21bklo, Gaia21ews & 276.66158 & 15.70446 & III & SN Ia & 0.056 &  & \cite{2021TNSCR3859....1B} \\
ZTF21acioiha & SN\,2021acgq & ATLAS21bkvo & 117.5581 & 72.5341 & II & SN Ia & 0.098 &  & \cite{2021TNSCR3748....1P} \\
ZTF21acipdhn & SN\,2021abzk & PS21msm & 157.55136 & 78.80644 & II & SN Ia & 0.079 &  & \cite{2021TNSCR3716....1S} \\
ZTF21acirwxt & AT\,2021acbf & ATLAS21bkoq & 335.56448 & 36.06381 & II &  &  &  &  \\
ZTF21acistjw & AT\,2021accn & ATLAS21blfb, PS21lvj & 8.94413 & -0.30567 & V &  &  &  &  \\
ZTF21aciuhyy & SN\,2021acgl & ATLAS21bkzw & 5.11115 & 29.44932 & III & SN Ia & 0.097 &  & \cite{2021TNSCR3766....1S} \\
ZTF21aciuque & AT\,2021acgm (also AT\,2013lh) & ATLAS21bkqh & 320.63354 & 15.41891 & VI &  &  &  &  \\
ZTF21acjbcdp & AT\,2021acsg & ATLAS21blay & 322.59706 & 3.14409 & VI &  &  &  &  \\
ZTF21acjbcdp & AT\,2021acsg & ATLAS21blay & 322.59705 & 3.14409 & VI &  &  &  &  \\
ZTF21acjbeap & AT\,2021acob & ATLAS21blcu & 334.34494 & 8.24779 & VI &  &  &  &  \\
ZTF21acjbedo & SN\,2021acpx & ATLAS21bkzt & 334.36867 & 13.38465 & II & SN Ia & 0.067 &  & \cite{2021TNSCR3736....1P} \\
ZTF21acjdwfe & SN\,2021acof &  & 345.46786 & 36.82734 & III & SN Ia & 0.107 &  & \cite{2021TNSCR3736....1P} \\
ZTF21acjnzdh & SN\,2021acoz & ATLAS21blvk & 102.10201 & 77.86272 & II & SN Ia & 0.095 &  & \cite{2021TNSCR3922....1P} \\
ZTF21acjosgg & SN\,2021acnf &  & 180.5743 & 64.71934 & I & SN Ia & 0.1058 &  & \cite{2021TNSCR4017....1G} \\
ZTF21acjrrle & AT\,2021acvr &  & 34.69133 & -16.66564 & VIII &  &  &  &  \\
ZTF21ackbhmm & SN\,2021adhg &  & 322.26635 & -12.85072 & II & SN Ia & 0.082 &  & \cite{2021TNSCR4017....1G} \\
ZTF21ackmeft & AT\,2021adfc &  & 316.26739 & -21.43709 & VI &  &  &  &  \\
ZTF21ackqvyg & AT\,2021adee &  & 45.64063 & -14.41236 & V &  &  &  &  \\
ZTF21ackthuy & AT\,2021adni &  & 56.49016 & -4.39208 & V &  &  &  &  \\
ZTF21ackyxaa & AT\,2021adqx & PS21lzj & 7.3939 & 2.80337 & VIII &  &  &  &  \\
ZTF21ackzlds & AT\,2021adjb &  & 359.80772 & -8.53838 & VI &  &  &  &  \\
ZTF21acldmwy & SN\,2021aefa & ATLAS21bmgf, PS21mjg & 87.02766 & -17.53591 & VIII & SN Ia & 0.11 &  & \cite{2021TNSAN.297....1R,2021TNSCR4027....1R} \\
ZTF21aclidzb & AT\,2021adij & PS22fk & 160.42135 & 25.57288 & II &  &  &  &  \\
ZTF21aclknsx & SN\,2021admz &  & 305.87894 & -23.44238 & II & SN IIn & 0.056 &  & \cite{2021TNSCR3952....1G} \\
ZTF21aclrkgs & SN\,2021admm &  & 13.2952 & -25.15131 & II & SN Ia-CSM & 0.108515 &  & \cite{2021TNSCR4071....1N} \\
ZTF21aclzvgz & SN\,2021adyo & ATLAS21bmxn & 174.22039 & 75.71104 & III & SN II & 0.0415 &  & \cite{2021TNSCR4064....1N} \\
ZTF21acmrhta & AT\,2021adwp &  & 273.89771 & 12.66109 & V &  &  &  &  \\
ZTF21acmtqzi & SN\,2021aebc & ATLAS21bmge, PS21mwv & 36.51753 & 31.20972 & III & SN Ia & 0.057785 &  & \cite{2021TNSCR3923....1B} \\
ZTF21acoruyj & SN\,2021aexp & ATLAS21bmox, PS21mrd & 24.5498 & -29.70116 & III & SN Ia & 0.07 &  & \cite{2021TNSAN.297....1R,2021TNSCR4027....1R} \\
ZTF21acouakb & AT\,2021aeke & ATLAS21bmxg & 134.73665 & 59.7688 & III &  &  & \tablenotemark{c} &  \\
ZTF21acphsgr & SN\,2021aeqi & ATLAS21bmoj & 330.70721 & -13.80123 & II & SN Ia & 0.052 &  & \cite{2021TNSCR4018....1N} \\
ZTF21acpocmd & SN\,2021aesp & ATLAS21bmpo, Gaia21ffo & 147.25064 & 39.23999 & II & SN Ia & 0.0411409996 &  & \cite{2021TNSCR3999....1S} \\
ZTF21acpzade & AT\,2021afjy &  & 355.31269 & -6.63483 & II &  &  &  &  \\
ZTF21acravtt & SN\,2021afzz &  & 179.06486 & -6.17787 & II & SN Ia & 0.083725 &  & \cite{2021TNSCR4145....1B} \\
ZTF22aaagqyx & SN\,2022cma &  & 73.58878 & -14.33213 & II & SN Ia & 0.056 &  & \cite{2022TNSCR.474....1K,2022TNSAN..46....1K} \\
ZTF22aaagrex & SN\,2022ccg & ATLAS22gdf, Gaia22arg, PS22blq & 43.12134 & 11.10408 & VII & SN Ia & 0.07 &  & \cite{2022TNSCR.542....1P,2022TNSAN..37....1B} \\
ZTF22aaahbrf & SN\,2022cdh & ATLAS22gat, PS22bmo & 56.64761 & 12.98938 & II & SN Ia & 0.035 &  & \cite{2022TNSCR.501....1P} \\
ZTF22aaahtqz & AT\,2022bdw & ATLAS22dth, Gaia22baj, PS22avi & 126.29315 & 18.58266 & II & TDE & 0.03782 &  & \cite{2022TNSCR.511....1A,2022TNSAN..50....1A} \\
ZTF22aaahuly & SN\,2022cca & ATLAS22ges, PS22bqe & 137.29314 & -3.86594 & II & SN Ic-BL & 0.042 &  & \cite{2022TNSAN..48....1A,2022TNSCR.500....1K} \\
ZTF22aaajlje & AT\,2022amc & ATLAS22cts, PS22bfb & 168.80077 & 7.9683 & VI & Other &  &  & \cite{2022TNSCR.291....1A,2022TNSAN..29....1A} \\
ZTF22aaajuiz & SN\,2022cfs & ATLAS22frs, PS22cms & 255.39057 & 32.95254 & II & SN Ia & 0.091 &  & \cite{2022TNSCR.457....1S} \\
ZTF22aaaoyme & SN\,2022cob & ATLAS22gcc, PS22bvc & 224.89748 & 14.12884 & III & SN Ic-BL & 0.0445 &  & \cite{2022TNSCR.509....1P} \\
ZTF22aaaqhgc & AT\,2022cms & ATLAS22gef & 58.49762 & 16.43121 & VI &  &  &  &  \\
ZTF22aaaqwar & SN\,2022coz & ATLAS22get, PS22brv & 121.23503 & 5.06504 & III & SN Ia & 0.035 &  & \cite{2022TNSCR.475....1P} \\
ZTF22aabfojs & AT\,2022crg & ATLAS22hfm, PS22drp & 217.16948 & 16.92929 & V &  &  &  &  \\
ZTF22aabimec & AT\,2022csn & ATLAS22ggz, Gaia22ayp, PS22bju & 112.22887 & 26.89703 & II & TDE & 0.148 &  & \cite{2022TNSCR.667....1S,2022TNSAN..61....1S,2022TNSCR3660....1A} \\
ZTF22aabjpii & SN\,2022cvp & ATLAS22gww, PS22drn & 221.12711 & 18.73436 & II & SN II & 0.044 &  & \cite{2022TNSCR.474....1K,2022TNSAN..46....1K,2022TNSCR.646....1B} \\
ZTF22aaboeyu & AT\,2022dlx & ATLAS22hob & 214.78224 & 12.65903 & VI &  &  &  &  \\
ZTF22aabsemf & AT\,2022dgt & ATLAS22hgi, PS22cji & 181.78881 & 62.27814 & VI &  &  &  &  \\
ZTF22aabsnrg & SN\,2022dfp & ATLAS22heg, PS22bwp & 128.81234 & 30.56737 & I & SN Ia & 0.07532 &  & \cite{2022TNSCR.600....1B} \\
ZTF22aabtqob & SN\,2022dfl & ATLAS22hcn, PS22ccq & 127.98901 & 3.01631 & III & SN Ia & 0.088 &  & \cite{2022TNSCR.843....1G} \\
ZTF22aabtyxu & AT\,2022eku & ATLAS22kxs, Gaia22byt & 167.50144 & 8.07103 & VII &  &  &  &  \\
ZTF22aabuamq & AT\,2022dsw &  & 171.99702 & 35.77912 & IX & AGN & 0.074 &  & \cite{2022TNSCR1070....1P} \\
ZTF22aabwemz & SN\,2022dkw & ATLAS22hmu & 218.95957 & 24.68283 & III & SN IIn & 0.036 &  & \cite{2022TNSCR.763....1M,2022TNSAN..74....1M} \\
ZTF22aabwyco & SN\,2022doi & ATLAS22hta & 226.91814 & 21.37483 & V & SN Ia & 0.092 &  & \cite{2022TNSAN..66....1M,2022TNSCR.700....1M} \\
ZTF22aacabjk & AT\,2022dsp & ATLAS22hyn & 256.4644 & 23.61804 & V &  &  &  &  \\
ZTF22aacinau & SN\,2022ebd & ATLAS22ino & 237.39294 & 13.9994 & II & SN Ia & 0.069875 &  & \cite{2022TNSCR.719....1B} \\
ZTF22aacmgor & SN\,2022duv & ATLAS22ifq, PS22bta & 120.11265 & -3.38228 & II & SN Ia & 0.054 &  & \cite{2022TNSCR.658....1D} \\
ZTF22aadesjc & SN\,2022fnl & ATLAS22keu, Gaia22cdd, PS22fdk & 233.42703 & 43.74597 & VII & SN IIn & 0.1035 &  & \cite{2022TNSCR1276....1S,2022TNSCR1282....1S} \\
ZTF22aaetqzk & SN\,2022gzi & ATLAS22nuf, Gaia22cei, PS22ewd & 266.5202 & 42.27627 & VII & SN IIn & 0.089 &  & \cite{2022TNSCR1478....1P} \\
ZTF22aahdxdt & AT\,2022ipr &  & 181.94545 & 4.09447 & X &  &  &  &  \\
ZTF22aaidexf & SN\,2022jdv &  & 227.75364 & 5.97167 & VIII & SN Ia-91T-like & 0.075 &  & \cite{2022TNSCR1200....1D,2022TNSAN.102....1D} \\
ZTF22aaikbez & AT\,2022jbz & ATLAS22nwq & 346.46355 & 41.59842 & VII &  &  &  &  \\
ZTF22aaimgkp & AT\,2022jks & ATLAS22odt & 180.95259 & 15.02741 & VII &  &  &  &  \\
ZTF22aaitzvr & SN\,2022jdf & ATLAS22nks, PS22eoi & 229.61761 & 5.23966 & VII & SN II & 0.04 &  & \cite{2022TNSCR1253....1F} \\
ZTF22aajibhc & AT\,2022jnp & ATLAS22puz & 212.28997 & 41.50829 & VII &  &  &  &  \\
ZTF22aajidyk & SN\,2022jnn & ATLAS22ntt, PS22ejw & 211.73102 & -25.01963 & VII & SN Ia & 0.049 &  & \cite{2022TNSCR1387....1H} \\
ZTF22aajlruz & SN\,2022jsj & ATLAS22ohs, PS22efd & 219.66979 & 43.7476 & VII & SN Ia & 0.09 &  & \cite{2022TNSCR1432....1S} \\
ZTF22aajpgof & SN\,2022jsg & ATLAS22otg, PS22ehg & 234.94636 & -20.16468 & VII & SN Ia & 0.07 &  & \cite{2022TNSCR1445....1H} \\
ZTF22aajrdad & SN\,2022jtj & ATLAS22obw & 142.7719 & 20.0688 & VII & SN Ia & 0.070942 &  & \cite{2022TNSCR1518....1S} \\
ZTF22aajrrzz & SN\,2022jut & ATLAS22nxl, PS22ejn & 241.45843 & 17.47988 & VII & SN Ia & 0.034129 &  & \cite{2022TNSCR1321....1J} \\
ZTF22aajzfxb & AT\,2022kto & ATLAS22pea & 328.33521 & -5.23381 & VII &  &  &  &  \\
ZTF22aakaygk & SN\,2022kbm & ATLAS22okl, Gaia22cxp, PS22eje & 246.32633 & 9.02035 & VII & SN II & 0.03 &  & \cite{2022TNSCR1410....1S} \\
ZTF22aakdnme & AT\,2022ket & ATLAS22ods, Gaia22cgy, PS22ewg & 203.66917 & 22.89847 & VII &  &  & \tablenotemark{c} &  \\
ZTF22aakejxf & SN\,2022kbc &  & 181.19546 & 9.43827 & VII & SN Ia-91bg-like & 0.0687 &  & \cite{2022TNSAN.116....1C,2022TNSCR1409....1C} \\
ZTF22aakmtrc & SN\,2022kpa & ATLAS22otb, PS22epr & 217.21103 & 14.99661 & VII & SN Ia & 0.086 &  & \cite{2022TNSCR1605....1S} \\
ZTF22aakvcqb & AT\,2019aaon & ZTF19aamakuv & 183.86497 & -2.42975 & VII &  &  &  &  \\
ZTF22aalmgkb & AT\,2022liy &  & 202.51259 & -19.40763 & VII &  &  &  &  \\
ZTF22aalxsmp & AT\,2022lrq & ATLAS22pun & 191.32769 & 13.49602 & VII &  &  &  &  \\
ZTF22aamcwwm & AT\,2022lni & ATLAS22pxv & 318.09583 & -6.15508 & VII &  &  &  &  \\
ZTF22aamrkns & SN\,2022kye & ATLAS22qhm, Gaia22cod, PS22epo & 315.44949 & -19.53634 & VII & SN II & 0.041 &  & \cite{2022TNSCR1882....1S}
\enddata

\tablenotetext{a}{Not clear if it is indeed rising at discovery.}
\tablenotetext{b}{Lower than the expected signal-to-noise ratio in the spectrum attempt.}
\tablenotetext{c}{Faded before the spectrum was attempted (could have been rapidly evolving or an artifact).}
\tablecomments{Since some queries are subsets of other queries, here we list only the most stringent query that produced each event.}

\end{deluxetable}
\end{longrotatetable}

\end{document}